\documentclass[fleqn,usenatbib]{mnras}

\usepackage{graphicx}	
\usepackage{amsmath,amssymb}	
\usepackage[usenames,dvipsnames]{xcolor} 
\usepackage[labelfont=bf]{subcaption} 
\usepackage{tabularx} 
\usepackage{relsize}
\usepackage{comment}
\usepackage{float}

\newcommand\Tstrut{\rule{0pt}{2.6ex}}         
\newcommand\Bstrut{\rule[-0.9ex]{0pt}{0pt}}   

\newcommand{\msun}{\,\mathrm{M}_{\odot}}

\newcommand{\gcmc}{\,\mathrm{g}\,\mathrm{cm}^{-3}}
\newcommand{\myr}{\,\mathrm{Myr}}
\newcommand{\myrs}{\,\mathrm{Myrs}}
\newcommand{\cmc}{\,\mathrm{cm}^{-3}}
\newcommand{\cms}{\,\mathrm{cm}^{-2}}
\newcommand{\kelvin}{\,\mathrm{K}}
\newcommand{\pc}{\,\mathrm{pc}}
\newcommand{\pcc}{\,\mathrm{pc}^3}
\newcommand{\au}{\,\mathrm{AU}}
\newcommand{\second}{\,\mathrm{s}}

\newcommand{\rangeto}{\,\text{ -- }\,}

\newcommand{\arepo}{\textsc{Arepo}}
\newcommand{\treecol}{\textsc{TreeCol}}
\newcommand{\disperse}{\textsc{DisPerSE}}
\newcommand{\fiesta}{\textsc{Fiesta}}

\usepackage[T1]{fontenc}
\usepackage{ae,aecompl}
\usepackage{newtxtext,newtxmath}

\title[Decaying turbulence in molecular clouds]{Decaying turbulence in molecular clouds: how does it affect filament networks and star formation?}

\author[Dhandha et al.]{
Jiten Dhandha,$^{1,2,3}$\thanks{E-mail: \href{mailto:jvd29@cam.ac.uk}{jvd29@cam.ac.uk}}
Zoe Faes,$^{1,4}$\thanks{E-mail: \href{mailto:zoe.faes@esa.int}{zoe.faes@esa.int}}
Rowan J. Smith$^{1,5}$\thanks{E-mail: \href{mailto:rjs22@st-andrews.ac.uk}{rjs22@st-andrews.ac.uk}}
\\
$^{1}$Jodrell Bank Centre for Astrophysics, Department of Physics and Astronomy, The University of Manchester, Manchester, M13 9PL, UK \\
$^{2}$Institute of Astronomy, University of Cambridge, Madingley Road, Cambridge, CB3 0HA, UK \\
$^{3}$Kavli Institute for Cosmology, Madingley Road, Cambridge, CB3 0HA, UK \\
$^{4}$European Space Agency, ESAC, Camino Bajo del Castillo s/n, Urb. Villafranca del Castillo, 28692 Villanueva de la Cañada, Madrid, Spain \\
$^{5}$SUPA School of Physics and Astronomy, University of St Andrews, North Haugh, St Andrews, Fife, KY17 9SS, UK
}

\date{Accepted XXX. Received YYY; in original form ZZZ}

\pubyear{2023}

\begin{document}
\label{firstpage}
\pagerange{\pageref{firstpage}--\pageref{lastpage}}
\maketitle

\begin{abstract}
The fragmentation of gas to form stars in molecular clouds is intrinsically linked to the turbulence within them. These internal motions are set at the birth of the cloud and may vary with galactic environment and as the cloud evolves. In this paper, we introduce a new suite of 15 high-resolution 3D molecular cloud simulations using the moving mesh code \arepo\, to investigate the role of different decaying turbulent modes (mixed, compressive and solenoidal) and virial ratios on the evolution of a $10^4\msun$ molecular cloud. We find that diffuse regions maintain a strong relic of the initial turbulent mode, whereas the initial gravitational potential dominates dense regions. Solenoidal seeded models thus give rise to a diffuse cloud with filament-like morphology, and an excess of brown dwarf mass fragments. Compressive seeded models have an early onset of star-formation, centrally condensed morphologies and a higher accretion rate, along with overbound clouds. 3D filaments identified using \disperse\, and analyzed through a new Python toolkit we develop and make publicly available with this work called \fiesta, show no clear trend in lengths, masses and densities between initial turbulent modes. Overbound clouds, however, produce more filaments and thus have more mass in filaments. The hubs formed by converging filaments are found to favour star-formation, with surprisingly similar mass distributions independent of the number of filaments connecting the hub.
\end{abstract}





\section{Introduction}
\label{s:Introduction}

The study of star formation spans several centuries, but the basic physical processes governing the earliest stages still remain poorly understood. The current paradigm of present-day star-formation suggests that the cold, dense phase of the interstellar medium (ISM), in the form of molecular clouds, undergo runaway gravitational collapse and birth stars in the process. As early as 1907, it was hypothesised that these clouds were not homogeneous, but contained substructure in the form of filaments connecting dense cores \citep{barnard1907nebulous}. Over the decades, the literature has vastly expanded to include modelling and characterisation of these sub-structures --- from protostars, to dense cores and clumps, to parsec-scale filaments, to several hundred parsec-scale molecular clouds. In particular, the discovery of the omni-presence of these filaments in the ISM by $\textit{Herschel}$ (e.g. \citealt{menshchikov2010filaments}) has brought a flurry of excitement and research on connecting the ISM geometry to the complex multi-scale physics of star formation, especially that of massive stars \citep[see][and references therein]{hacar2022initial}.

The typical characteristics of molecular clouds are generally understood: they are dense ($\gtrsim\text{100}\cmc$), and largely composed of H$_2$, HI, H$^+$, CO, and other molecules of C, H, O and N \citep{bergin1997chemical}. Efficient line cooling by CO and a high population of absorptive dust grains impart a characteristically low temperature of $10\kelvin\rangeto20\kelvin$ \citep[see, e.g., reviews][]{BerginTafalla2007,dobbs2014formation} --- conditions that are perfect for gravitational collapse of the cloud. The collapse is, however, not linear; the clouds are permeated by supersonic, compressible turbulence, as demonstrated in the seminal work by \citet{larson1981turbulence}, which play a crucial role in the evolution of the cloud \citep[see, e.g., reviews][]{mac2004control,mckee2007theory}. Fully characterising the collapse of molecular clouds is a hard task due to the complexity of the scales at which they occur --- from kiloparsec-scale in the galactic interstellar medium, down to less than 0.01 parsec for single proto-stellar objects. Idealised simulations reveal that the gas develops a lognormal density probability function (PDF) which relates to the properties of the turbulence and possibly the subsequent stellar initial mass function (IMF) \citep[e.g.][]{klessen2000pdf,krumholz2005general,burkhart2018star,federrath2013star}. Generally, turbulence is seen as playing a dual role: it supports the cloud against global collapse, and also accelerates it through local compressions which give rise to a complex filamentary networks of dense gas.

The mass, size and density of filaments can span several orders of magnitude, with lengths ranging from of $0.01\pc$ to several hundreds of pc, and masses from $0.01\msun\rangeto10^{5}\msun$ \citep{hacar2022initial}. They generally have central number densities higher than $1000\cmc$. These filaments fragment into denser clumps and cores, where stars and stellar clusters are born \citep{hartmann2002flows,andre2010filamentary,myers2011filamentary,hacar2013cores,konyves2015census} --- like $\textit{beads on a string}$. This makes filaments a key ingredient in any star formation theory, especially in the post-$\textit{Herschel}$ era. Numerical simulations have shown that contracting sheet-like clouds accrete material onto filaments with velocity flows perpendicular to them; and once channelled onto the filament, material flows longitudinally towards cores \citep[e.g.][]{smith2011quantification,gomez2014filaments,smith2016nature}. This is not just inferred using simulations, but also seen in observations \citep{kirk2013filamentary,peretto2013global,hacar2017gravitational,liu2021kinematics}.
Hence, filaments form the bridge for collapsing gas to be transferred from large scales into dense cores where stars may form. Modern theoretical works have explored the many different mechanisms that give rise to these filaments \citep{hennebelle2013origin,federrath2016universality,smith2016nature,abe2021classification}. 

Despite their importance, there is no universal, physically motivated definition for a filament in both observations and numerical simulations. Early pioneering works used an idealised filament geometry of an infinite isothermal cylinder \citep{ostriker1964equilibrium,ostriker1964oscillations,larson1985cloud,inutsuka1992self}, and found there existed a critical-line mass above which it becomes gravitationally unstable and fragments into cores. However, modern studies find filaments to be far from ideal, with complex substructure. They are dynamic objects affected by magnetic fields \citep{nagasawa1987gravitational,nakamura1993fragmentation,adam2016planck,ade2016planck,li2019magnetized}, external pressures \citep{fiege2000helical}, turbulence \citep{hartmann2002flows}, filament shape \citep{gritschneder2017oscillating}, and evolve through interactions with their environment \citep{smith2014nature}.
There has been significant discourse on the possibility of a universal filament width of $0.1\pc$, first suggested by \citet{arzoumanian2011characterizing}. 
This remains contested with some suggesting that it may be an effect of the methodology used to derive the width \citep{smith2014nature} or an effect of distance and resolution \citep{Panopoulou22} . For a full discussion on filaments, see the review by \citet{hacar2022initial}.

The evolution of filaments is inextricably linked to that of massive stars.
Massive stars play a key role in galaxy evolution as they regulate the gas-star cycle through wind and radiation, and eventual supernovae explosions \citep{mac2004control}. Studies of massive star formation are also important to constrain the IMF and characterise star-formation rates in galaxies. Despite this, a full understanding of their formation remains elusive. Two modern theories exist: core accretion and competitive accretion. In the core accretion model, massive self-gravitating pre-stellar cores that condense from a fragmenting clump (protostellar region) give rise to massive stars \citep{shu1987star,McKee03}. In the competitive accretion model, the formation of dense clumps occurs simultaneously with stellar clusters. Initially, this theory proposed that stars grew by Bondi-Hoyle type accretion \citep{Zinnecker82,bonnell2001competitive}. However modern versions of the theory focus on accretion during the large scale collapse of a parent clump of gas whereby protostars at the centre of a potential accrete mass from large radii and become massive over time \citep{bonnell2006star,smith2009simultaneous}. For a comprehensive review on this topic we refer to \citet{tan2014massive} and references therein. Regardless of the mechanism, high-resolution observations from ALMA have hinted at a relation between complex networks of filaments and massive-star forming clumps \citep[e.g.][]{peretto2013global,hacar2018alma, Motte18review,Chen19,Dewangan20}. Naturally, intersections of filaments, or `hubs', are the primary suspect for these clumps, having very high densities that can support massive star formation \citep{myers2009filamentary,peretto2012pipe,schneider2012cluster,kirk2013filamentary,rayner2017far}.

The complex dynamics of turbulence and global collapse in molecular clouds, along with the formation and evolution of filamentary networks and hubs, form the perfect basis for numerical simulations — to constrain the $\textit{recipe for star-formation}$. The use of magnetohydrodynamical (MHD) codes is widespread for this purpose  \citep{fryxell2000flash,Springel2010}. The theory, and subsequently simulations, are divided into two competing scenarios on this topic: the turbulent scenario and the global hierarchical and chaotic collapse scenario. 

In the turbulent scenario, molecular clouds are in virial equilibrium between self-gravity and strongly supersonic turbulence and evolve over many free-fall times, maintaining a low rate of star formation throughout \citep[see reviews by][and references therein]{mac2004control,BallesterosParedesi2007,mckee2007theory,BerginTafalla2007,gnedin2016star}. In order for this scenario to play out, there must be a continuous source of turbulence that supports these clouds. Proposed mechanisms are (i) internal feedback from stellar sources \citep[e.g.][]{norman1980clumpy,franco1983self,mckee1989photoionization,krumholz2006global,goldbaum2011global}, or (ii) external driving by converging flows in the warm neutral medium \citep[e.g.][]{vazquez2007molecular,vazquez2010molecular,heitsch2008rapid,klessen2010accretion,micic2013cloud}. However, the former fails to explain the non-thermal velocities observed in clouds without star-formation, while the latter drivers are too weak to support the cloud. Simulations of this scenario drive turbulence at large scales \citep[e.g.][]{federrath2008density,heyer2004universality,federrath2016universality}, akin to `shaking' the simulation box periodically (called turbulent forcing).
	
The opposing view of global hierarchical collapse proposes that molecular clouds are transient, irregular features born from large-scale HI flows \citep[see review ][and references therein]{vazquez2019global}. Turbulence is generated by the converging flows and amplified by the collapse itself (so-called `chaotic infall') giving rise to a moderately supersonic cloud. Most of the turbulent energy decays through a large number of weak shocks \citep{smith2000shock,smith2000distribution}, and the evolution of the cloud takes place in the form of a mass cascade --- collapse occurs on all scales with smaller scales accreting from their parental structure \citep[e.g.][]{Hopkins13}. Small, dense regions contract later, but on shorter timescales than large, diffuse ones. Overall, the global contraction onto local centres of collapse occurs on $\sim$ one free-fall timescale. The complex inner structure of molecular clouds is explained by the amplification of density fluctuations through self-gravity, magneto-hydrodynamic instabilities \citep{heitsch2005formation, heitsch2006birth,heitsch2007magnetized} and thermal instabilities \citep{hennebelle1999dynamical,audit2005thermal}. As transient objects, the gravitational potential energy of the cloud starts low and increases monotonically until it catches up with the kinetic and thermal energies. Thus, rather than evolving in true virial equilibrium, the cloud evolves far from it until gravity dominates the energy balance, causing the near equipartition condition $E_\text{grav} \sim 2E_\text{kin}$ \citep{vazquez2007molecular,bonilla2022gravity,vazquez2019global}. Due to the successes of this model, it forms the basis of our numerical simulations. This scenario can be simulated by feeding a turbulent velocity field to the molecular cloud, and letting self-gravity take over the evolution of the cloud leading to a `decay' in the initial conditions.

Depending on the source, the mode of the turbulence in clouds can be compressive (curl-free) or solenoidal (divergence-free). Dynamical mechanisms (galactic spiral shocks and converging flows) or stellar feedback (supernovae and radiation-pressure-driven shells) excite compressive modes in clouds, while high shear environments (at galactic centres, for example) or magneto-rotational instabilities induce solenoidal modes \citep{federrath2016link}. 

In this work, through a suite of 15 gravoturbulent simulations, we explore how freely decaying turbulence seeded with different initial turbulent modes (mixed, compressive and solenoidal) and dynamical states (characterised by their virial ratios) affect the evolution of a molecular cloud. The paper is broadly divided into two main parts. In Section~\ref{s:methods}, we outline the numerical model and methodology  used to identify and characterise filamentary structures in the evolved clouds. In Section~\ref{s:results}, we present and discuss an array of results ranging from cloud morphologies and density distributions, to star-formation statistics, IMFs and filament statistics. Finally, we outline some caveats and future work in Section~\ref{s:caveats} and summarise our conclusions in Section~\ref{s:conclusions}.


\section{Methods}
\label{s:methods}


\subsection{Numerical model}

We perform our simulations on the moving mesh code \arepo\, \citep{Springel2010,Weinberger2020}, a multi-physics cosmological (M)HD code that 
uses a finite volume approach to solve the MHD equations, and performs calculations on an unstructured mesh formed by the Voronoi tessellation of a set of points that move freely with the fluid flow. This results in a Galilean-invariant solution accurate to second order in all four dimensions. The freedom of geometry allows for complex dynamics to be simulated. The smooth fluid flow automatically adjusts resolution at localised clusters (i.e., higher resolution at high density regions), while the ability to perform adaptive mesh refinement on a grid captures the effects of shocks and fluid instabilities well. Furthermore, all calculations are made on a hierarchical time-step that is optimised for parallelisation. In our setup, the refinement of the Voronoi mesh is set such that it applies Jeans refinement criteria of 8 cells per local Jeans length to regions with number density $n > 100\cmc$, to avoid artificial fragmentation \citep{Truelove97}. Furthermore, to avoid discontinuous jumps in the cell size, adjacent cells may not differ in their radii by a factor larger than two throughout the simulation volume. The self-gravity of the gas is calculated using the standard \arepo\, gravitational tree \citep{Springel2010}. The effect of magnetic fields and stellar feedback processes such as supernovae explosions, stellar winds, jets and outflows are absent in our models.

The chemical network in the simulations consist of five species: H$_2$, HI, H$^+$, CO and C$^+$, with the inclusion of dust grains. The hydrogen network is based on \citet{GloverMacLow2007a,GloverMacLow2007b} and the CO network is that of \citet{NL1997}. The treatment includes formation and destruction of molecules through collisions and chemical reactions, and ionisation by interstellar radiation field (ISRF) and cosmic rays. The strength of the ultraviolet ISRF field is set to $G_0 = 1.7\text{\,Habing units}$, as calculated by \citet{draine1978photoelectric} assuming same strength and spectrum of the field as the solar neighbourhood. The cosmic ray ionisation rate is set to $\xi_\text{H} = 3\times10^{-17}\second^{-1}$ for atomic hydrogen, as calculated using observations of 7 protostars by \citet{van2000limits}, and twice the amount for molecular hydrogen. This model was first implemented in \arepo\, by \citet{smith2014co}. The effect of ISRF attenuation due to $\text{H}_2$ self-shielding, $\text{CO}$ self-shielding, $\text{CO}$ shielding by $\text{H}_2$ and dust absorption is modelled using the \treecol\, algorithm developed by \citet{TreeCol2012}. Heating and cooling of the gas from radiative processes is computed alongside the chemistry using the atomic and molecular cooling function described in \citet{clark2019tracing}.


\subsection{Star formation characterisation}

We characterise star formation in the simulations through the framework of sinks, first introduced in \arepo\, by \citet{greif2011simulations}. We use a modified version detailed in \citet{tress2020simulations}, briefly discussed here. A sink is formed when a gas cell passes a density threshold of $\rho_\text{sink}$ (equivalently number density of $n_\text{sink}$) within an accretion radius $r_\text{acc}$. Once a cell or group of cells passes this threshold, the code checks if the region is indeed gravitationally bound --- i.e. potential energy $U > 2\left(E_\text{kinetic} + E_\text{thermal}\right)$ --- and has a negative divergence of velocity and acceleration to ensure gas is unambiguously collapsing. On passing the checks, the cells are replaced with collisionless `sink' particles which inherit the combined mass and velocity of the parent gas cells and are located at the centre of mass of the region. The sink particles henceforth interact with their environment solely through gravity and may accrete mass (above the density threshold) from adjacent gas cells. We set $\rho_\text{sink}$ to $\approx 4 \times 10^{-16}\gcmc$ (or $ n_\text{sink} \approx 1.7 \times 10^8\cmc$), which is orders of magnitude higher than the average molecular cloud density, and close to those seen in protostellar regions \citep[e.g.][]{whitworth2001empirical}. To ensure the entire instability is contained within the sink particle, the sink accretion radius $r_\text{acc}$ is set to 110\% of the Jeans length at $n_\text{sink}$, yielding a value of $\approx 200\au$. At this length scale, sinks may represent individual stellar systems (in case of low mass sinks) or small stellar sub-clusters (in case of high mass sinks). 

Due to the absence of complex feedback processes like stellar winds and jets, sinks may be better thought of as `accretion reservoirs' --- they do not represent the mass in stars, but the mass available for the formation of stars. In actuality, a variable percentage of the sink mass would go towards the star formation, due to varying efficiencies depending on stellar environments and length scales.



\subsection{Initial conditions}

\begin{figure}
    \centering
    \includegraphics[width=0.9\linewidth]{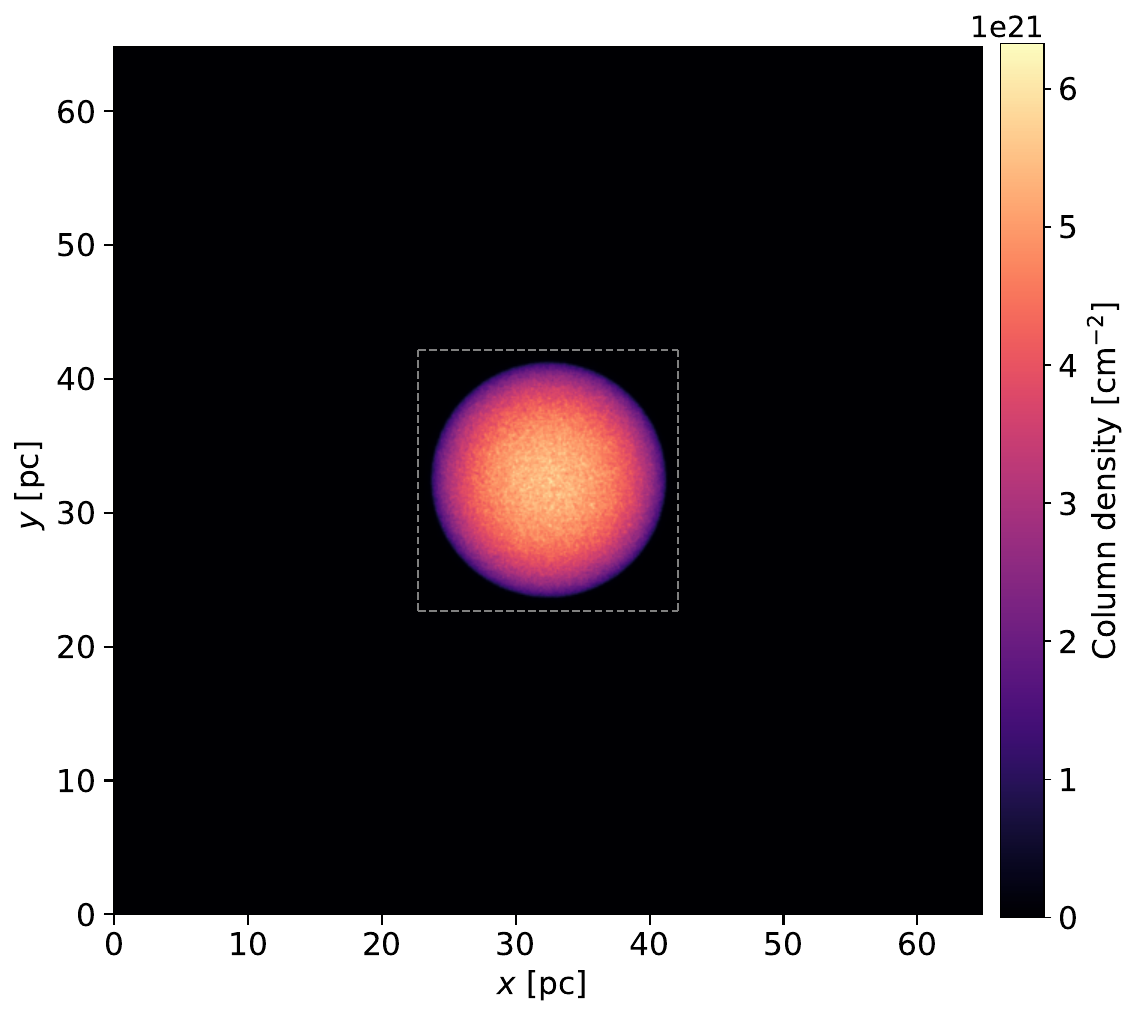}
    \caption{Column density projection in the $z$-direction of our initial condition for the molecular clouds. The cloud is spherically symmetric and centred in the simulation box. The region enclosed by dashed grey lines, spanning $\approx 19.4\pcc$ delineates the region of interest (ROI). Our analysis is restricted to this region throughout the paper.}
    \label{fig:initialcloud}
\end{figure}

\begin{figure*}
	\includegraphics[width=\textwidth]{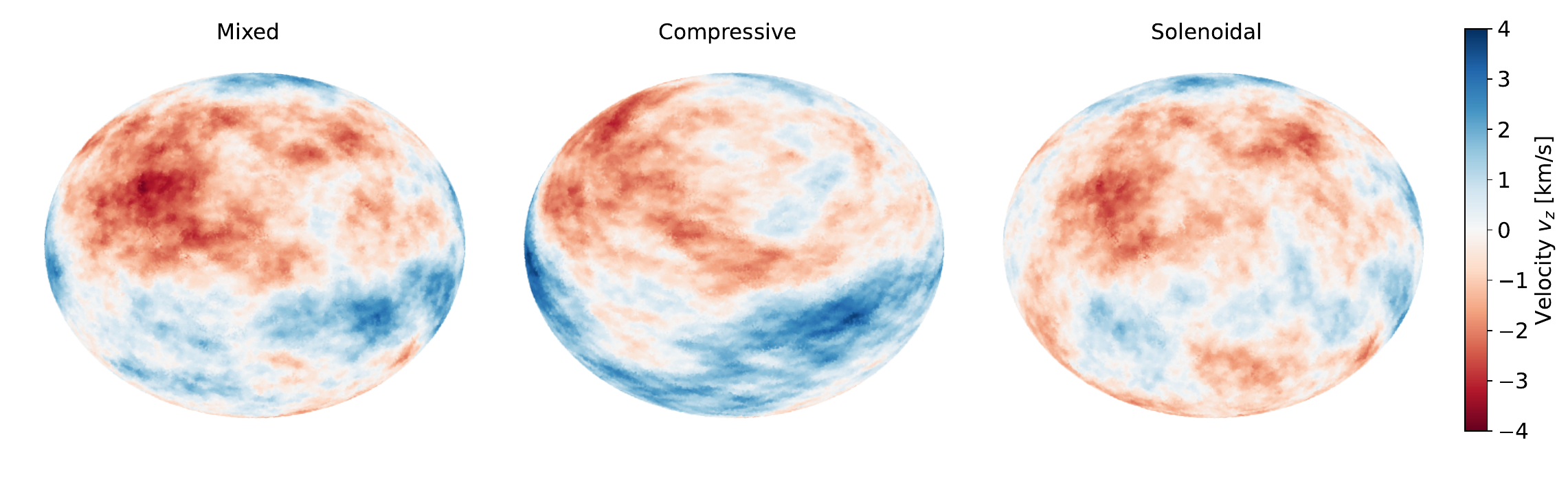}
	\caption{Example of turbulent velocity field with virial-type equipartition, generated using RNG seed = 57. Figures show the heat-map of $v_z$ component of the velocity field for mixed, compressive and solenoidal modes respectively. The velocities shown are on the surface on the cloud, for a portion of the cloud facing the viewer; the clouds are indeed spherical, but appear ellipsoidal due to skewed axes. Note the large-scale nature of fluctuations and in particular, the striated pattern typical of compressive turbulence.}
	 \label{F:velocitymaps}
\end{figure*}

\begin{table*}
\caption{\small Key simulation parameters for the suite of 15 simulations performed. For each turbulent mode, three seeds (26, 57 and 90) were used to generate turbulent velocity fields. Once generated, they are scaled to either virial balance (simulations 1 -- 9) or non-virial ratios (simulations 10 -- 15). The virial ratios $\alpha_\text{vir} = 2.0$ and $\alpha_\text{vir} = 0.6$ correspond to underbound and overbound clouds respectively. A Mach number $\mathcal{M} \approx 3\rangeto4$, defined as the ratio of velocity dispersion and sound speed $\sigma_v/c_s$, is typical of the GHCC model \citep{vazquez2019global}. As expected, the overbound and underbound Mach numbers are $\sqrt{\alpha_\text{vir}}$ scaling of their virial counterparts.} 
\label{T:InitialConditionsSimulations}
\resizebox{\textwidth}{!}{%
\begin{tabular}{cccccc|cccccc}
\hline
\hline
 & Simulation & \begin{tabular}[c]{@{}c@{}}Turbulence\\ mode\end{tabular} & \begin{tabular}[c]{@{}c@{}}RNG\\ seed\end{tabular} & \begin{tabular}[c]{@{}c@{}}virial ratio\\ $\alpha_\text{vir}$\end{tabular} & \begin{tabular}[c]{@{}c@{}}Mach number\\ $\mathcal{M}$\end{tabular} & & Simulation &  \begin{tabular}[c]{@{}c@{}}Turbulence\\ mode\end{tabular} & \begin{tabular}[c]{@{}c@{}}RNG\\ seed\end{tabular} & \begin{tabular}[c]{@{}c@{}}virial ratio\\ $\alpha_\text{vir}$\end{tabular} & \begin{tabular}[c]{@{}c@{}}Mach number\\ $\mathcal{M\Tstrut\Bstrut}$\end{tabular}\\ \hline 
1  & M26 & Mixed         & 26         & 1.0      & 4.01  &  10 & M57U & Mixed         & 57         & 2.0      & 5.82  \\
2  & M57 & Mixed         & 57         & 1.0      & 4.12  &  11 & M57O & Mixed         & 57         & 0.6      & 3.19  \\
3  & M90 & Mixed         & 90         & 1.0      & 4.04  &  12 & C57U & Compressive   & 57         & 2.0      & 5.66  \\
4  & C26 & Compressive   & 26         & 1.0      & 4.07  &  13 & C57O & Compressive   & 57         & 0.6      & 3.10  \\
5  & C57 & Compressive   & 57         & 1.0      & 4.00  &  14 & S57U & Solenoidal    & 57         & 2.0      & 5.85  \\
6  & C90 & Compressive   & 90         & 1.0      & 4.10  &  15 & S57O & Solenoidal    & 57         & 0.6      & 3.20  \\
7  & S26 & Solenoidal    & 26         & 1.0      & 4.09  &     &      &               &            &          &   \\
8  & S57 & Solenoidal    & 57         & 1.0      & 4.14  &     &      &               &            &          &   \\
9  & S90 & Solenoidal    & 90         & 1.0      & 4.07  &     &      &               &            &          &   \\
\hline                                                                    
\end{tabular}
}
\end{table*}

The initial molecular cloud used across all simulations is a spherical, uniform cloud of mass $10^4\msun$ and radius $8.83\pc$, embedded centrally in a periodic bounding cubic box of side $64.82\pc$, as shown in Figure \ref{fig:initialcloud}. The large simulation box ensures that no boundary issues affect the evolution of the gas cloud, thus simulating an isolated, dense, molecular cloud in warm, tenuous ISM. The simulation initially consists of 4.01 million cells, of which 4 million belong to the cloud and 10,000 to the surrounding ISM, which changes as the mesh evolves. The cloud has a number density of $\approx 120\cmc$ and temperature of $27\kelvin$. The surrounding ISM environment has temperature of more than $7500\kelvin$ and a much lower number density of $10^{-5}\rangeto10^{-4}\cmc$. The global free-fall time of the cloud is $t_\text{ff} \approx 6.5\myrs$.

The cloud is initially composed entirely of atomic hydrogen with a helium abundance (by number) of $0.1$, H$^+$ abundance of $0.01$, carbon abundance of $1. 4\times 10^{-4}$, oxygen abundance of $3.2\times 10^{-4}$ and a metallic abundance of $1.0\times10^{-7}$, all relative to hydrogen abundance. The dust-to-gas ratio is $1/100$. The gas cells in the cloud are yet to be assigned velocities, which is discussed in the next section.


\subsection{Initial turbulence}

The turbulence used in these simulations is initialized on large scales and decays freely (through small-scale shocks), similar to \citet{bate2005origin}, \citet{clark2005onset}, \citet{GloverMacLow2007b}, \citet{vazquez2007molecular}, \citet{bate2009dependence}, \citet{Girichidis11} to name a few.

We generate homogeneous, normalised turbulent velocity fields using the code described in \citet{Girichidis11}.
The algorithm creates a Fourier ($\textbf{k}$-space) grid and calculates the energies from an energy spectrum $E(k)$, which can be set to either Kolmogorov spectrum $E(k) \sim k^{-\frac{5}{3}}$ \citep{kolmogorov1941local} or Burger's turbulence $E(k) \sim k^{-2}$ \citep{burgers1939mathematical}. We choose the latter since it is consistent with observations and simulations of supersonic, compressible turbulence in molecular clouds \citep[][albeit centred around turbulent forcing]{kritsuk2007statistics,schmidt2009numerical,federrath2013universality}. The code then generates amplitudes and random phases for each $\textbf{k}$-space grid point. The amplitude is just the energy $E(k)$, while the phase is produced through random number generation (RNG) on a uniform distribution from a specified seed parameter. This gives $\hat{\textbf{u}}(\textbf{k})$, the Fourier transform of the turbulent velocity field $\textbf{u}(\textbf{x})$. The field is optionally Helmholtz-decomposed into longitudinal (compressive) and transversal (solenoidal) fields, or left unchanged (mixed turbulence, which is 2 parts compressive to 1 part solenoidal) depending on the mode of turbulence chosen. A final inverse-Fourier transform gives the turbulent velocity field $\textbf{u}(\textbf{x})$, which is then normalised before outputting. The velocity field generated is defined on a Euclidean grid of size $N^3$ where $N$ is the number of grid points along an axis and can be varied as a parameter.

To simulate a decaying turbulent velocity field, we feed the output generated by this code to the molecular cloud at the start of the simulation. Since these are generated on a cubic grid of dimensions $N^3$ and the geometry of \arepo\, is a Voronoi mesh, the velocities have to be interpolated and scaled to fit the spherical molecular cloud. The $N^3$ velocity grid is first centred on the cloud, and all gas cells belonging to the molecular cloud in \arepo's ROI are assigned a velocity by linear interpolation between the cubic grid points. The computational intensity of interpolation sets the limit on how finely the velocities are resolved, i.e. how big $N$ is. In our setup, we use $N=750$ which corresponds to a length of $\approx 0.025\pc$ in the simulation.

The velocities $v_i$, once assigned to the cloud, are then scaled in accordance to the virial ratio:
\begin{equation}
	 \label{E:virialcondition}
		\alpha_\text{vir} \equiv \dfrac{2E_\text{kin,tot}}{|E_\text{grav}|} \approx \left(\sum_{i}^{N_\text{cells}}\mathbf{v}_i^2\right)\bigg/\left(\dfrac{3}{5}\dfrac{GM}{R}\right)
\end{equation}
where $i$ iterates over the cloud cells in the simulation, $M$ is the total mass of the cloud, $R$ is its radius, and $G$ is the gravitational constant. For virial-type equipartition, the cloud must be $\alpha_\text{vir}=1$. Since the gravitational potential energy of the molecular cloud in our simulation remains unchanged, i.e. the cloud mass is not varied, an under-bound cloud is one which has large velocities leading to $\alpha_\text{vir}>1$, while an over-bound cloud is one which has a weak velocity field, i.e. $\alpha_\text{vir}<1$. Figure \ref{F:velocitymaps} shows the $v_z$ component of virial balanced clouds for different turbulent modes.

Having discussed the simulation technique, we now present the suite of fifteen different simulations with varying turbulent velocity modes and virial ratios. Nine of these simulations explore the statistics of different turbulent velocity modes at virial-type equipartiation by using different random seeds, while the other six explore different virial ratios for the three modes. Table \ref{T:InitialConditionsSimulations} summarises the different initial velocity conditions. These simulations are all available publicly on Zenodo\footnote{See Section~\ref{sec:Data Availability} for data availability.}.


\subsection{Filament identification}
\label{ssec:Filament identification}

\begin{figure}
\centering
\includegraphics[width=0.98\linewidth]{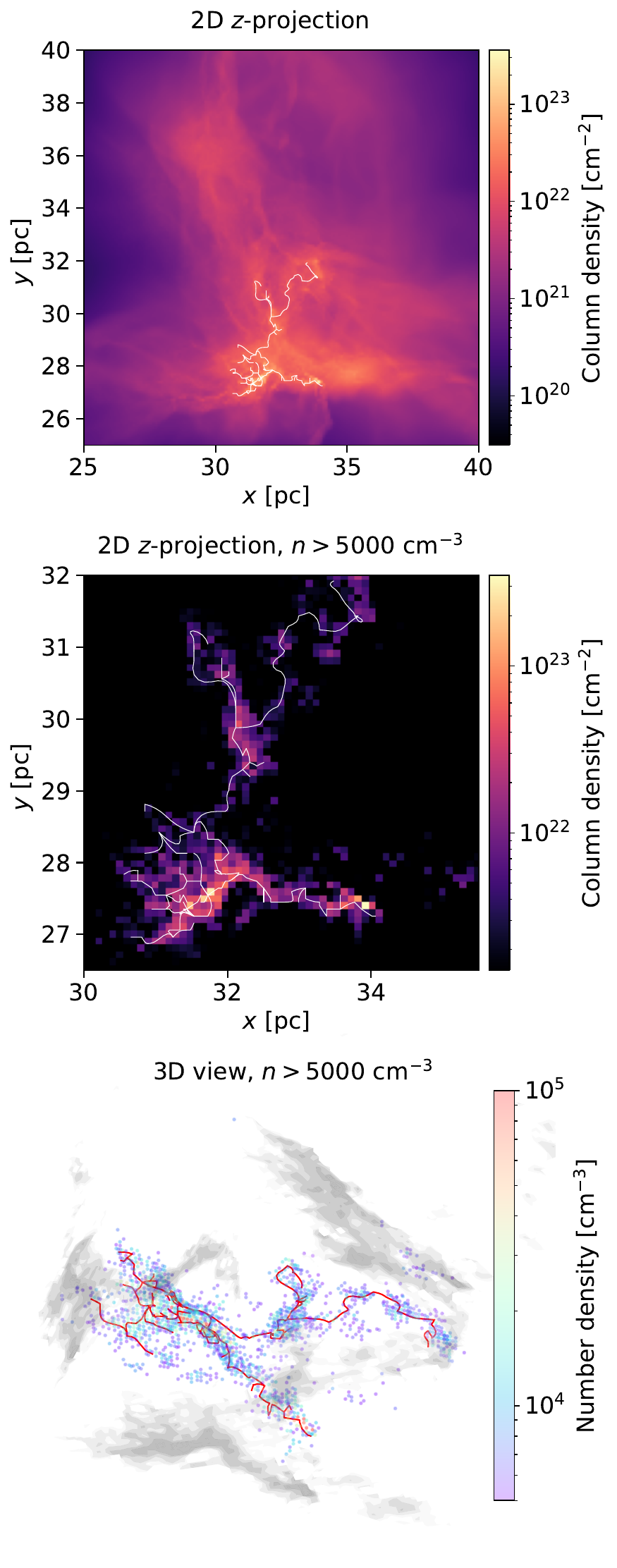}
\caption{Example of filamentary structures identified using \disperse\, in simulation C57. Top panel shows a projection of the entire central region of the simulation box with filaments (white). Middle panel is similar but shows the projection of dense regions  $>5000\cmc$ only. Bottom panel shows a 3D plot of these dense regions (colour), their axial projections (grey shadows), and filaments (red).}
\label{fig:FilamentExample}
\end{figure}

\begin{figure*}
\centering
\includegraphics[width=\linewidth]{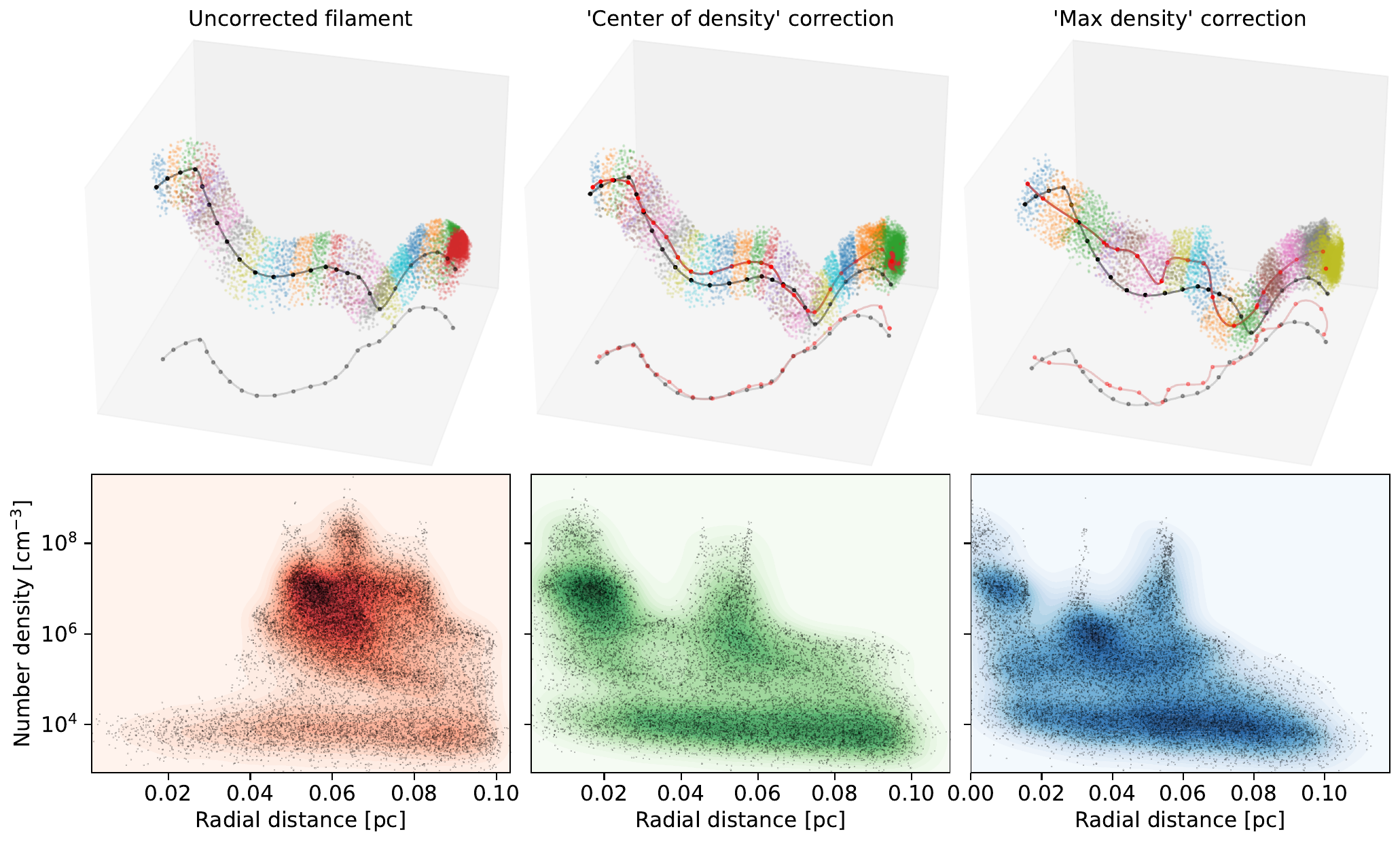}
\caption{Example of the two filament spine correction algorithms implemented --- `centre of density' and `max density' --- for a randomly chosen filament, and their corresponding radial density profile. \textbf{Top row:} 3D plots showing the filament (and projection) pre-correction in dark grey (light grey) and post-correction in red (light red), with the \arepo\, cells (colour) belonging to each longitudnal cylinders of radius $R_\text{cyl}=0.1\pc$ along the filament spine. \textbf{Bottom row:} Plots shows the density of \arepo\, cells versus their distance from filament spine (black scatter), along with contours (colour).}
\label{fig:FilamentCharacterization}
\end{figure*}

As stated in the introduction, filaments are not strictly defined features. They can be diffuse, clumpy, dense, smooth, stub-like or long and tortuous, and often blend into their background. This makes it hard to assign a length, width or mass to them, whether in observational images or in simulations. However, their undebatable ubiquity at all scales in our Universe warrants a method of characterising them. In order to identify filamentary structures in our molecular clouds, we use the open-source tool \disperse\, \citep[DIScrete PERsistent Structures Extractor;][]{sousbie2011persistent}.\footnote{For more on filament identification codes, \citet{chira2018fragmentation} provides an excellent comparison of the several available tools.} Although the code was originally written for the identification of web-like structures in cosmology \cite[e.g.][]{kleiner2017evidence,laigle2018cosmos2015,kraljic2018galaxy,libeskind2018tracing}, the so-called `cosmic web', it has found wide use in molecular cloud studies for identifying filaments \cite[e.g.][]{arzoumanian2011characterizing,federrath2016universality,smith2016nature,smith2020cloud}. The algorithm employs Morse theory to identify structures like voids, sheets, filaments and peaks in an $n$-dimensional topological space. It first constructs a Morse-Smale complex of an input density distribution (or field), and identifies two key features in it: critical points --- points where the gradient of the distribution is zero, i.e. $\nabla\rho(x,y,z)=0$ --- and integral lines --- curves that are tangent to the gradient at every point, and start and end at critical points. The critical points may be maxima, minima or saddle-points. The set of integral lines that join maxima to saddle points, called ascending 1-manifolds, delineate filaments in the input field, and form the `skeleton' network of filaments. Each filament is defined by a number of sampling points it passes through, and ends in critical points.

Since \disperse\, requires uniform cubic grids for identifying filaments, the density distribution of \arepo\, simulations within the ROI are cast from Voronoi geometry to a uniform cubic grid of size $N^3$ using \arepo's in-built ray-tracing algorithm. Due to the computational intensity of running \disperse\, on 3D density cubes, we use $N=200$ which sets an artificial resolution limit of $\approx 0.1\pc$ for our filament networks. Thus, in the hierarchical substructures of filaments, our access is limited to a specific lowest scale. This is not unlike observations, where the resolution limit is set by the telescope.

$\disperse$\, has several free parameters for identifying filaments. The persistence threshold parameter is the most crucial and largely determines what the identified network looks like. A pair of critical points, joined by a filament, is defined to have a numerical value called `persistence'. This quantity has significance in defining the robustness of a topological feature, and is the main focus of the persistence theory. An intuitive explanation for this is offered in \citet[][Section 4]{sousbie2011persistent}, which we briefly summarise here. Consider a function used to model the density of a region. The relative `height' of the adjacent local maxima and minima in the density field contains information about how tenuous or significant that topological feature is. Formally, this is known as the lifetime of the component in the excursion set. Small relative heights are akin to `noise' (or small scale perturbations) in the local density field. The higher the threshold, the denser the regions being traced, while low values normally generate many spurious, short filaments.

Most often in observation-based literature, this value is provided in the context of filamentary structures in 2D column density maps, where it is set to $\approx10^{21}\cms$ \citep{arzoumanian2011characterizing,federrath2016universality}. In 3D simulations, \citet{chira2018fragmentation} have used two values, $100\cmc$ and $5000\cmc$, for their analysis. In this work, we pick $n_\text{thresh}=8500\cmc$ by analysing the regions traced for various persistence thresholds, as done in Figure~\ref{fig:FilamentExample}. A more detailed description of our methodology is provided in Appendix~\ref{appendix1}, where we also quantify the errors introduced by this parameter.  We also fix a threshold angle, set to $90^\circ$, above which longer filaments are split into shorter pieces.


\subsection{\textit{A \fiesta~of filaments}}
\label{ssec:fiesta}

In order to study the nature of these filaments in our simulations, we have developed a set of tools to work with outputs from \arepo\, and \disperse, and analyse them. This includes reading and writing data for both softwares, visualisations of the simulations and filaments, and also functionality for characterising their properties such as lengths, masses, density profiles and more, which we describe in the next section. These tools are released as an open-source Python toolkit called
\textbf{FI}lam\textbf{E}ntary \textbf{ST}ructure \textbf{A}nalysis (\fiesta), available on GitHub\footnote{\fiesta\, library: see documentation at \href{https://fiesta-astro.readthedocs.io}{fiesta-astro.readthedocs.io}.}. Although primarily used for the aforementioned software, the algorithms developed can be applied more generally to study any filamentary networks defined on structured grids or unstructured meshes.

\subsubsection{Filament characterisation}

The filament network identified by \disperse\, only returns the co-ordinates of the filament spines. The length, width, density profile, and mass have to be determined through other means. The conversion of a 1D spine to a 3D tortuous cylinder is not a trivial one.

The sampling points for each filament are first fit by a 1D curve passing through them, by means of cubic B-spline interpolation\footnote{performed using \textsc{scipy}\, \href{https://docs.scipy.org/doc/scipy/}{docs.scipy.org/doc/scipy} library's interpolation module.}, hence giving each filament a parameterised characteristic function that fully defines it: $\boldsymbol{f}(s) = (x,y,z)$. Here, $s\in[0,1]$ is the normalised distance of a point along the filament and $(x,y,z)$ are the co-ordinates of the same point. The utility of this function is that it allows the filament to be sampled as finely or coarsely as needed. The length is then trivially calculated by integrating the Euclidean length along the characteristic function.

Determining the mass is a much harder task since filaments are not rigidly defined structures, but diffuse and merge with their background. To assign mass, we use an algorithm which relies on splitting the filament into longitudinal cylinders and querying the mass of all \arepo\, cells within these cylinders. The step-wise algorithm is as follows: consecutive sampling points along a filament are joined by vectors. Each vector is rotated and translated to the origin, along with all \arepo\, cells in a cubic region surrounding the base of the vector. This region selection avoids re-centering and rotation of all cells in the simulation box, which would be too computationally heavy. A cylindrical region is then defined with a fixed radius of $R_\text{cyl}$ around the vector. In the origin-centred vector frame of reference, this constrains the $x\text{--}y$ values, while the $z$ axis is constrained by the ends of the vector. \arepo\, cells within this cylinder are queried and the process is repeated for all sampling points in the filament. The total mass contained in these \arepo\, cells is defined as the mass of the filament, correcting for double-counting between cylinders.

The top panel of Figure~\ref{fig:FilamentCharacterization} shows an example of the spine of the filament and the cylinders queried around it.

\subsubsection{Caveats to the characterisation}
\label{sss:caveatstocharacterization}
Filament identification by \disperse\, is not perfect --- the main issue lies with the $0.1\pc$ artificial resolution introduced by ray casting the finely resolved densities on the Voronoi mesh to a coarsely resolved cubic grid. The grid casting washes out local overdensities --- hence, irregular clumps on length scales shorter than the resolution are not resolved properly. Filaments passing through or near these dense clumps may not be identified reliably by \disperse, and are noted to have inflated masses. To partially mitigate this, we apply two corrections: first, all filaments shorter than $0.1\pc$, the minimum resolution of the cubic grid, are considered unreliable detections and discarded. Second, we apply a spine correction to the filament where the spine is re-adjusted using the resolved Voronoi mesh. This is done by querying all \arepo\, cells in a spherical region of radius $R_\text{sph}=0.1\pc$ around each sampling point of the filament, and then shifting the point to either the densest point in the sphere (`max density' method) or its mean (`centre of density' method). The former method introduces more tortuosity in the filament due to strong local density fluctuations in the resolved mesh, as seen in Figure~\ref{fig:FilamentCharacterization}. 
Hence, we adopt the `center of density' method in this paper. This gives smoother filaments, and doesnt bias the resultant filament completely towards dense cores. The mode of correction noticeably affects the filament profiles as discussed in the next section.

\subsubsection{On filament widths and profiles}

As prefaced in the introduction, filament widths have been intensely studied in literature. A `universal' width of $0.1\pc$ has been suggested by many \citep{arzoumanian2011characterizing,malinen2012profiling,juvela2012profiles,kirk2015role,federrath2016universality}, but remains contested \citep{smith2014nature,Panopoulou22}. In this work, we adopt a fixed $R_\text{cyl}=0.1\pc$ radius for each filament since it is the resolution of the grid used for \disperse, hence twice the suggested universal width. For the purpose of this work, a fixed radius can shed light on how much mass is contained in and around these filaments as a comparative metric, rather than a definitive property of filaments in these turbulent clouds.

In observational data or 2D simulation projections, a Plummer fit is often used for filament density profiles \citep[see, e.g., ][]{federrath2016universality} to calculate its width. We find that such a fit is not well suited for 3D filaments.
Figure~\ref{fig:FilamentCharacterization} shows an example of the density profile obtained for a randomly chosen filament, for both types of spine correction mentioned in section \ref{sss:caveatstocharacterization}. Indeed, there is a filament profile in the density, but it is embedded within a background of both diffuse and clumpy regions scattered across the profile. The sharp peaks are likely due to dense cores (as suggested by the magnitude of their density) offset from the filament spine. The hierarchical nature of filaments also means that there are mini-filaments (or `ribbons') that thread the main filament. Thus the profiles vary from filament to filament, with some less messy than others, making it hard to define a single fit that would work for all. A similar conclusion is reached in \citet{smith2014nature} where sub-filaments contribute to a non-smooth 3D density profile. In our case, there is an additional layer of subjectivity introduced by the form of spine correction, which affects the density profile as seen in Figure~\ref{fig:FilamentCharacterization}.

\section{Results}
\label{s:results}

\begin{figure}
\centering
    \includegraphics[width=1.0\linewidth=]{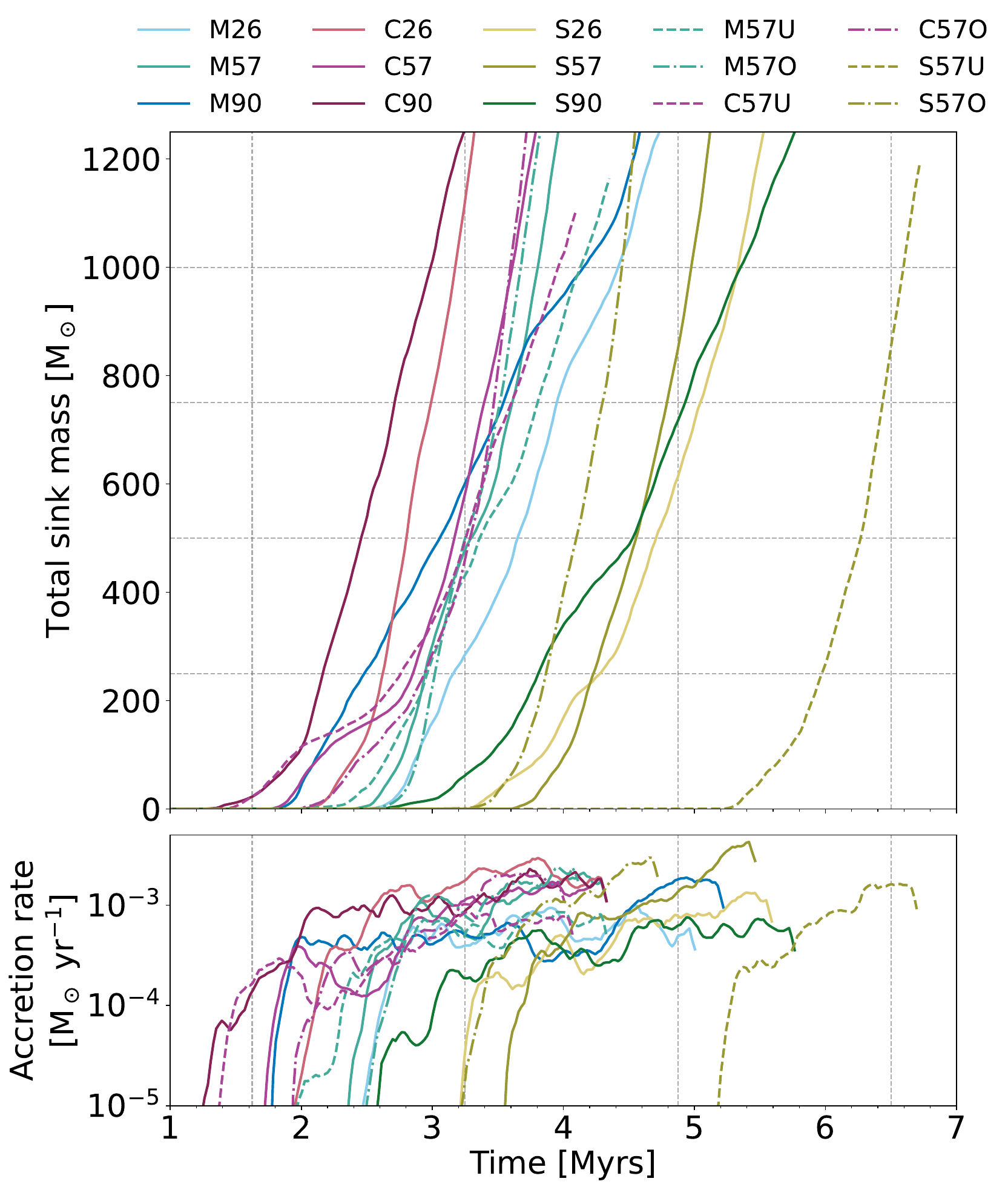}
    \caption{\textbf{Top:} Evolution of total sink mass as a function of time for all 15 simulations. The horizontal dashed grey lines demarcate the epochs of interest used in our analysis, corresponding to a fixed total sink mass --- $250 \msun$, $500 \msun$, $750 \msun$ and $1000 \msun$ --- achieved at different times for each simulation. The vertical grey lines show quarters of 1 free-fall timescale ($t_\text{ff}\approx6.5\myrs$). \textbf{Bottom:} Accretion rate onto sinks as a function of time.}
    \label{fig:sink_mass_evolution}
\end{figure}

\begin{figure*}
\centering
    \includegraphics[width=1.0\linewidth=]{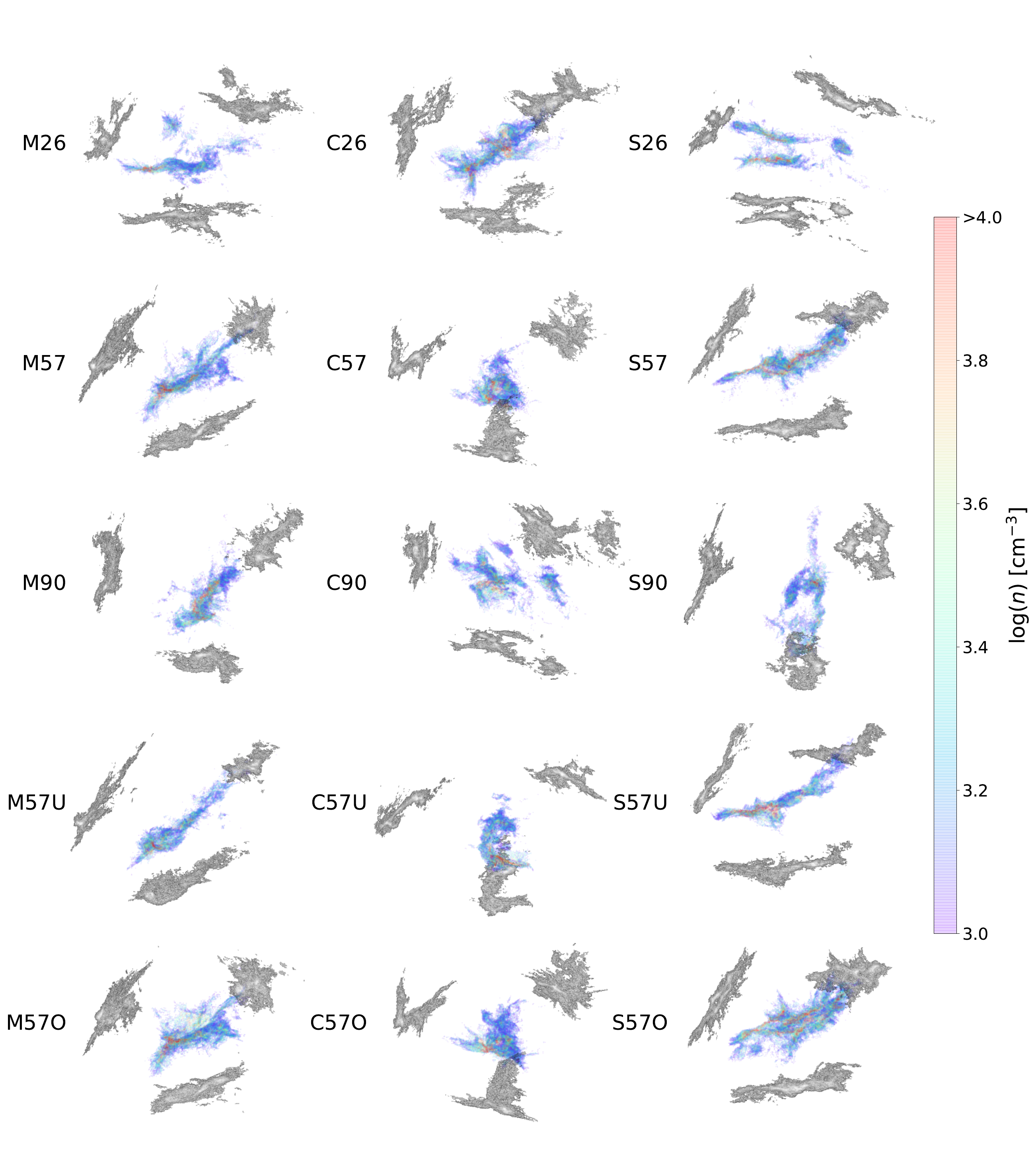}
    \caption{Morphology of all clouds at the last epoch, $M_\text{sink} = 1000\sun$. The 3D plots show dense regions $>1000\cmc$ (colour), and their axial projections (grey). Compressive seeded clouds show more compact centrally condensed structures, while solenoidal seeded clouds are more filament-like.} 
    \label{fig:AllClouds}
\end{figure*}

\begin{figure*}
    \includegraphics[width=\textwidth]{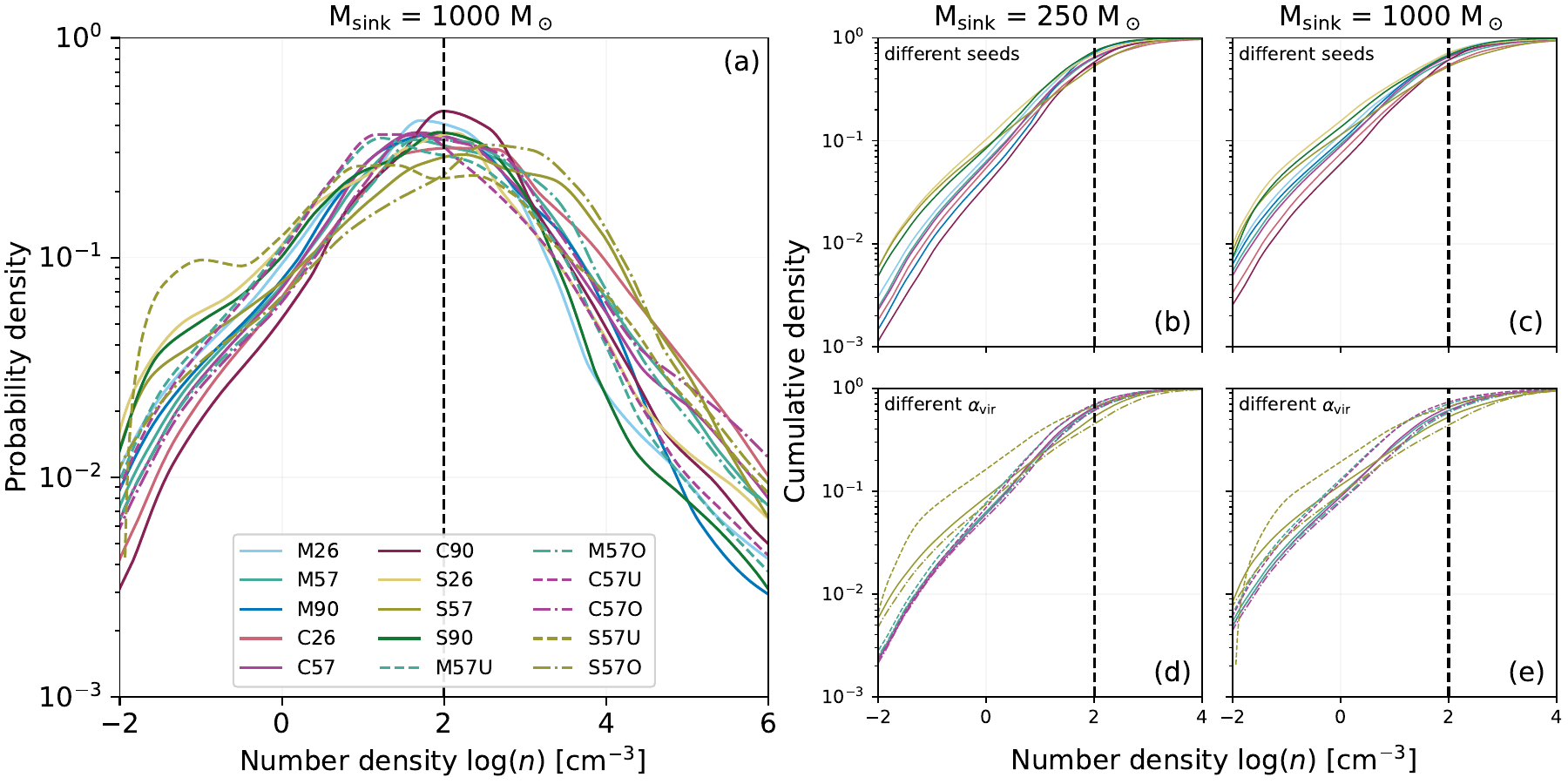}
    \caption{\textbf{(a)} Mass-weighted PDFs of number density for all simulations at the $M_\text{sink}=1000\msun$ epoch, and the mean $n \sim 100\cmc$ (dotted black line). \textbf{(b)} CDFs of number density for clouds in initial virial balance, but different seeds, at $M_\text{sink}=250\msun$ epoch. \textbf{(c)} Same as (b) but at $M_\text{sink}=1000\msun$ epoch. \textbf{(d)} CDFs of number density for clouds with different virial ratios at $M_\text{sink}=250\msun$ epoch. \textbf{(e)} Same as (d) but at $M_\text{sink}=1000\msun$ epoch.}
    \label{fig:PDFs}
\end{figure*}

\begin{figure}
    \includegraphics[width=\linewidth]{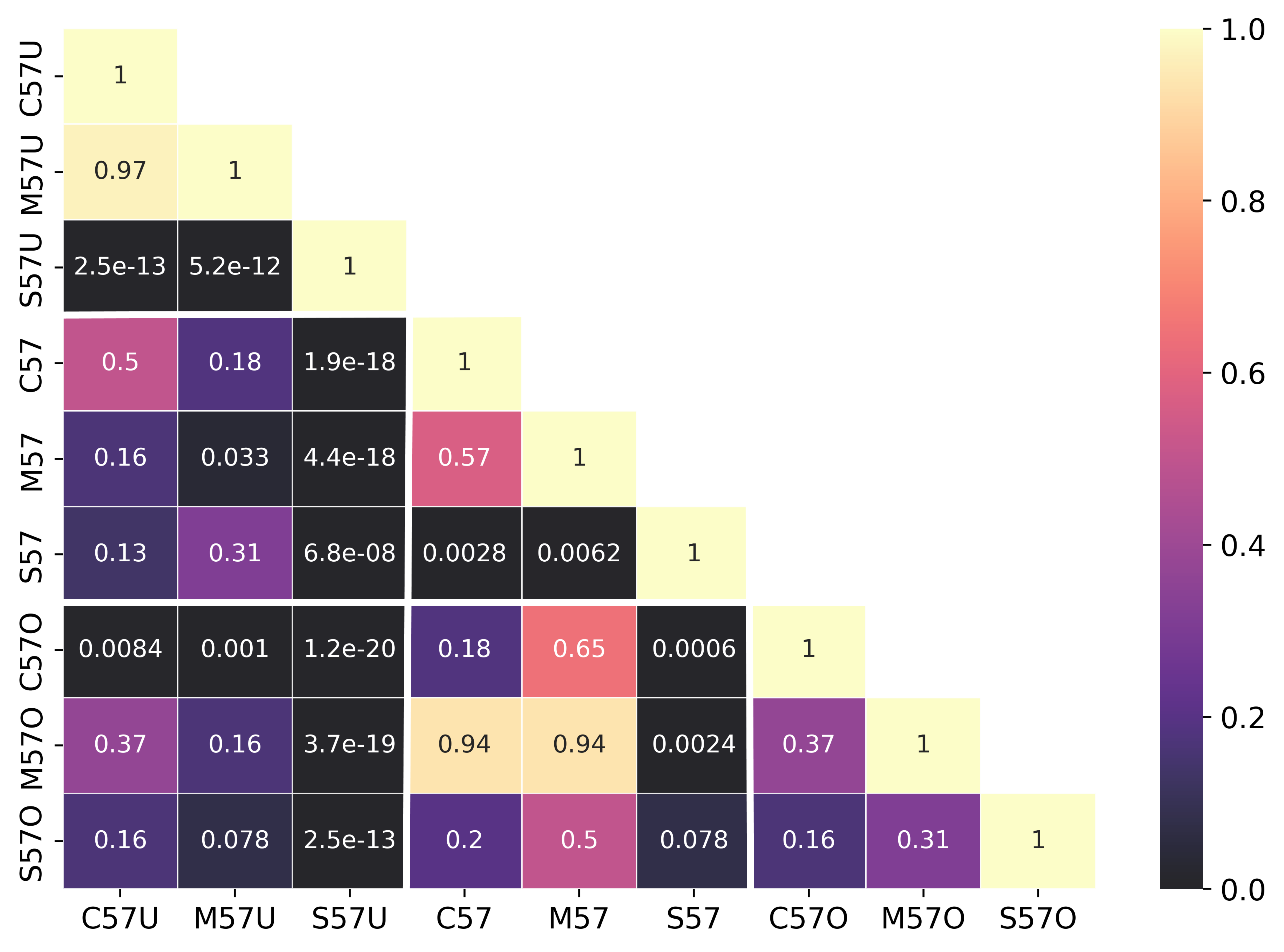}
    \caption{Two-sampled Kolmogorov-Smirnov (KS) test $p$-values for each combination of log-scaled mass-weighted CDFs in Figure~\ref{fig:PDFs}. We focus on a single RNG~seed~$=57$ to compare the effects of turbulent mode and boundedness.}
    \label{fig:KStests}
\end{figure}

\begin{figure*}
\centering
    \includegraphics[width=1.0\linewidth]{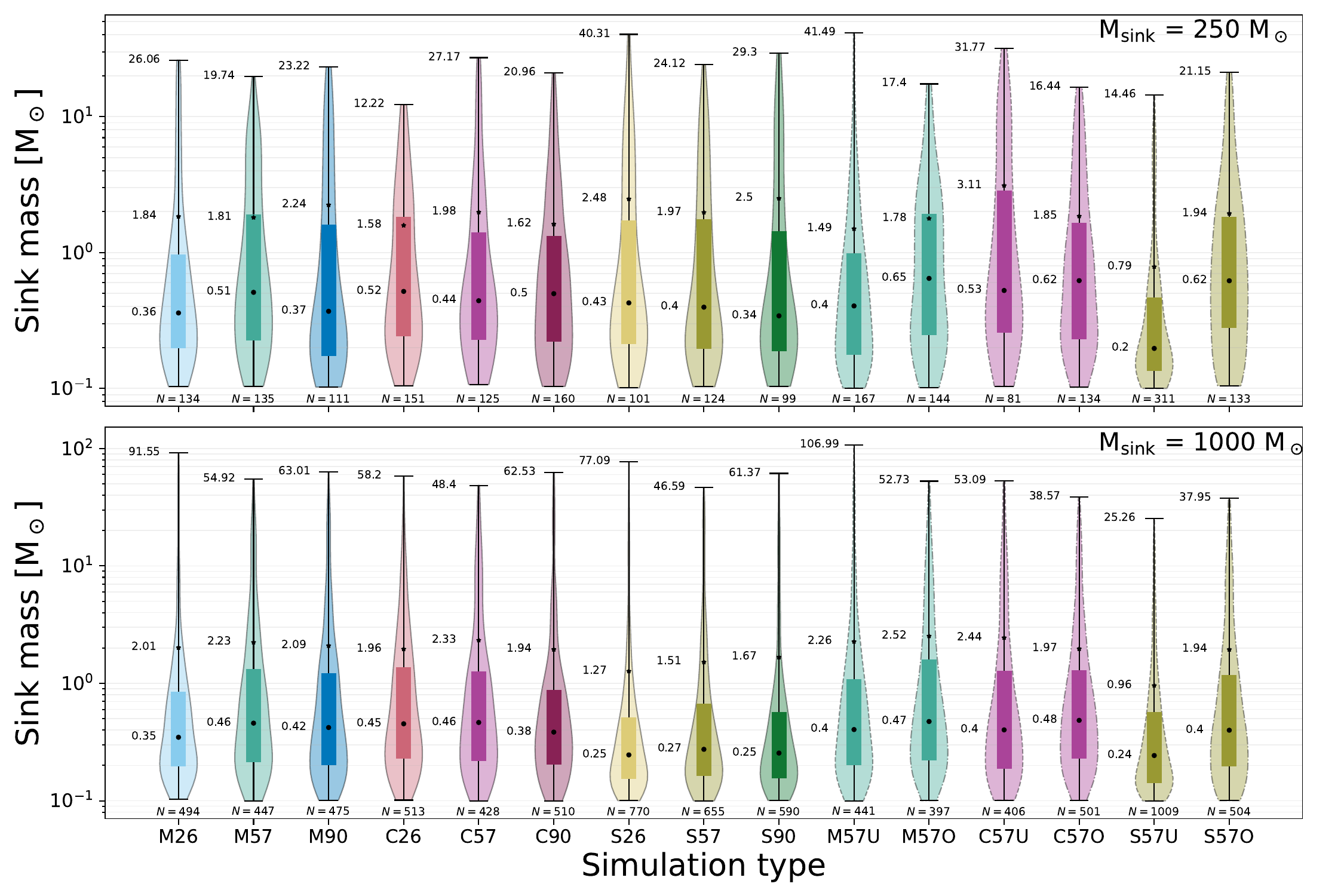}
    \caption{Violin plots of sink masses for all simulations at $M_\text{sink}=250\msun$ and $M_\text{sink}=1000\msun$ epochs, truncated to those $>0.1\msun$ since they are likely to form stars. The plots show 25\%--75\% quartiles (coloured bars), along with the median (black dot) and mean (black star), and total number of sinks along the $x$-axis.}
    \label{fig:Sink_mass_stats}
\end{figure*}

\begin{figure*}
    \begin{subfigure}[b]{0.49\textwidth}
    \centering
        \includegraphics[width=1.0\linewidth]{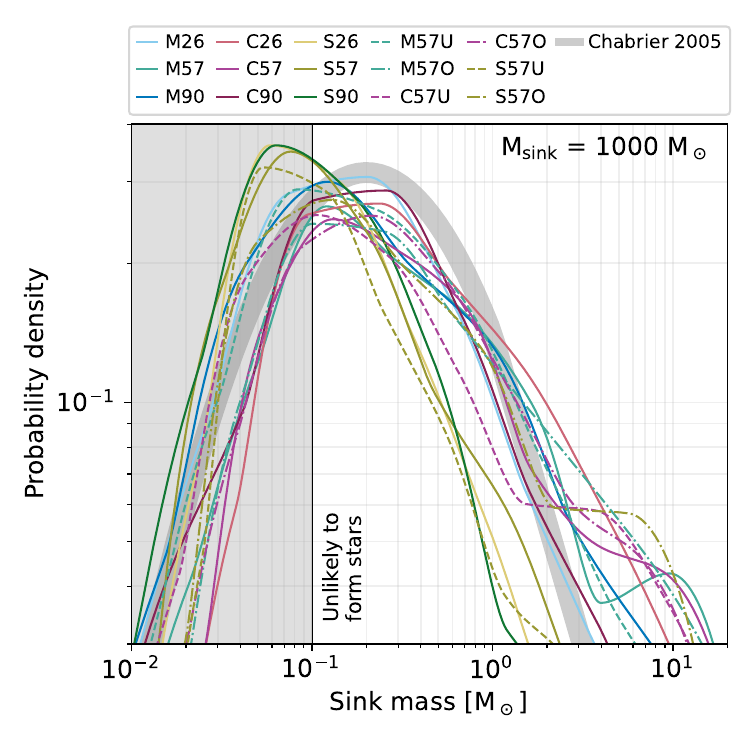}
        \caption{}
        \label{fig:Sink_mass_histogarm}
    \end{subfigure}
    \begin{subfigure}[b]{0.49\textwidth}
        \centering
        \includegraphics[width=1.0\linewidth]{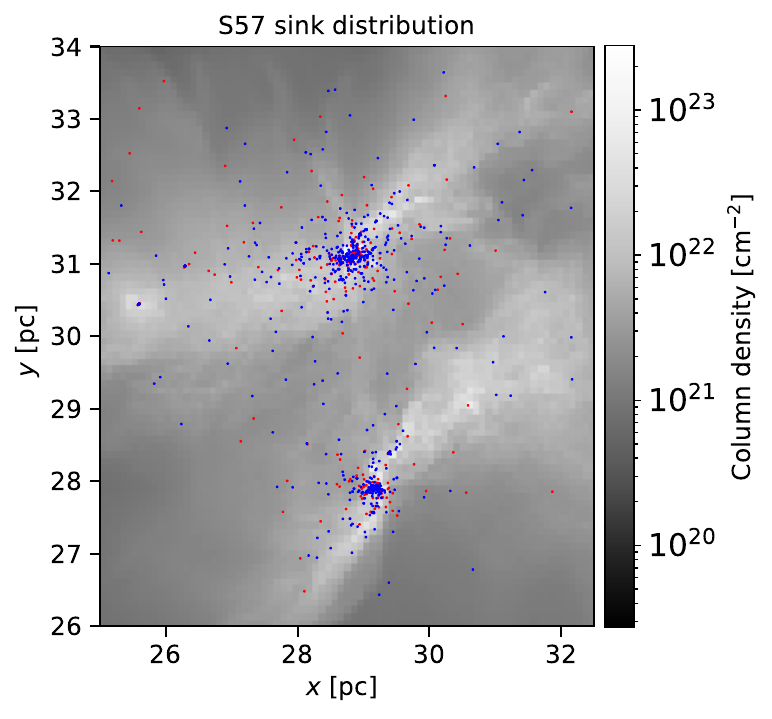}
        \caption{}
    \label{fig:BrownDwarfDistribution}
    \end{subfigure}
    \caption{\textbf{(a)} Probability density function of sink masses for all simulations at the final epoch, taking the form of an IMF (note, however, that these mass functions are for \textit{sinks}, and not \textit{stars}, so it is strictly not an IMF). The \citet{chabrier2005initial} model is generated using \href{https://github.com/keflavich/imf/}{github.com/keflavich/imf} and normalized to the displayed mass range. Most notably, the solenoidal clouds show an excess of low mass sinks (brown dwarfs) $<0.1\msun$. \textbf{(b)} 2D projection of simulation S57 (grey), overlayed with the distribution of low mass sinks $<0.1\msun$ (red scatter) and high mass sinks $>0.1\msun$ (blue scatter). They appear to be similarly clustered, indicating that these brown dwarfs are not forming in unique spatial environments.}
\end{figure*}

\begin{figure}
\centering
    \includegraphics[width=1.0\linewidth]{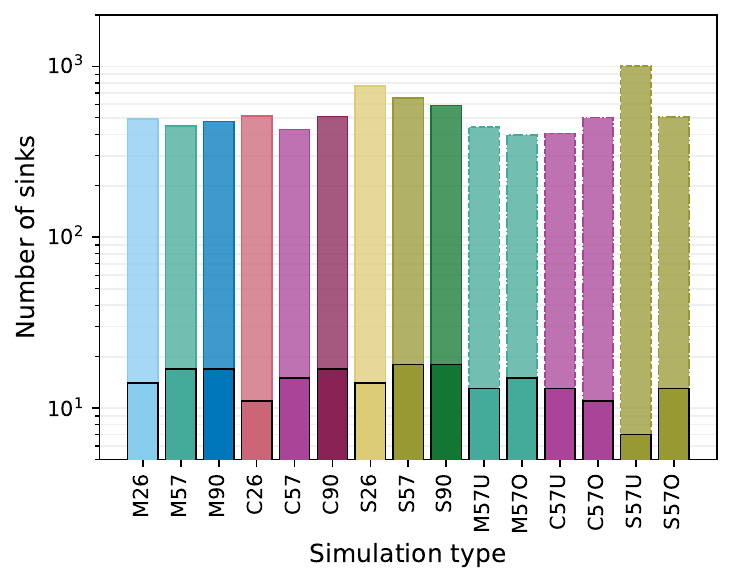}
    \caption{\textit{Which initial conditions give rise to more massive stars?} Plot shows histogram of sink count $>0.1\msun$ (light, with coloured outline) and $>16\msun$ (dark, with black outline) at $\mathrm{M_{sink}}=1000\msun$ epoch. Formation of massive sinks appears largely unaffected by the type of seeded turbulence, forming evenly across all simulations ($\sim15$ in count), since it is likely governed by local collapse dynamics. The relative abundance (normalized mass function) is shown in Figure~\ref{fig:Sink_mass_histogarm}.}
    \label{fig:Massive_sink_count}
\end{figure}


Throughout this work, we shall focus our analysis on four epochs of interest based on the total mass in sinks: $250 \msun$, $500 \msun$, $750 \msun$ and $1000 \msun$. Since the total mass of the molecular cloud is $10^4 \msun$, these epochs correspond to 2.5\%, 5\%, 7.5\% and 10\% global star-formation efficiency of the cloud (or more accurately, sink formation efficiency). As the cloud evolves differently in each simulation, these thresholds are achieved at different times. Hence, instead of comparing clouds at fixed time-steps, we compare them at similar stages in their evolutionary track. In the former approach, any differences in filamentary structures or density distributions between the simulations would be an artefact of the stage of cloud collapse; intuitively, one would expect compressive and overbound clouds to collapse faster than solenoidal and underbound clouds. The latter approach gives a truer picture of how the modes and virial ratios differ fundamentally once seeded into the cloud, and is closer to how an observer may interpret the cloud without knowledge of its absolute time evolution and only current star formation efficiency. Figure \ref{fig:sink_mass_evolution} shows the total sink mass evolution as a function of time, which we shall discuss in detail in Section \ref{ssec:sink mass}.


\subsection{Cloud morphology}

We first investigate the morphology of the clouds. Figure \ref{fig:AllClouds} shows a 3D view of the dense regions in each cloud ($>1000\mathrm{\,cm}^{-3}$) at the final epoch, along with their 2D projections. We see that despite being at the same stage of collapse (i.e., same mass in sinks), solenoidal seeded turbulence tends to form elongated `filament-like' clouds at the global scale, while compressive seeded turbulence forms `centrally condensed' clouds. Looking at the evolution\footnote{see Zenodo repository, available in Section~\ref{sec:Data Availability}, for animations.}, there is a definite rapid convergent flow in the compressive seeded clouds which forms an intermediate filament-like structure, before ending with a centrally condensed morphology. In comparison, solenoidal seeded clouds have a diffuse and slow collapse, decaying into compressive sub-modes through self-gravitation at much later times. There is a similar trend seen across underbound and overbound clouds, where the underbound models generate a more `filament-like' morphology, while overbound models generate a more `centrally condensed' morphology for the same initial turbulence seed. To quantify this observation better, we calculate the point cloud spread of all cells above $1000\mathrm{\,cm}^{-3}$ as the mass-weighted standard deviation:
\begin{equation}
    \sigma^2 = \dfrac{\mathlarger{\sum}_{i}^{n}m_i\left((x_i-\bar{x}^*)^2 + (y_i-\bar{y}^*)^2 + (z_i-\bar{z}^*)^2\right)}{\mathlarger{\sum}_{i}^{n}m_i}
    \label{equation:pointspread}
\end{equation}
where $(x_i,y_i,z_i)$ are the co-ordinates of cell $i$, $m_i$ is the mass of the cell, and $(\bar{x}^*,\bar{y}^*,\bar{z}^*)$ is the mass-weighted centroid of all cells. Table~\ref{table:pointspread} shows this quantity for all simulations. There is a clear large dispersion of dense gas in the solenoidal seeded clouds, while mixed turbulence and compressive seeded turbulence forms more compact clouds. This is also seen in the underbound versus overbound cases, where S57U seems to have the largest spread of any simulation, almost twice as much as some compressive clouds. This naturally raises the question, \textit{are the solenoidal clouds really diffuse and less dense than others?} --- a question readily answered by looking at the probability density functions (PDFs) of these clouds.

\begin{table}
\centering
\caption{Point cloud spread of cells above $1000\mathrm{\,cm}^{-3}$, as defined in Equation~\ref{equation:pointspread}, at the epoch $M_\text{sink}=1000\msun$.}
\begin{tabular}{cc|cc|cc}
    \hline
    \hline
    Sim. & $\sigma$ & Sim. & $\sigma$ & Sim. & $\sigma$ \\
    \hline
    M26 & 9.56\,pc & C26 & 8.48\,pc & S26 & 11.31\,pc \\
    M57 & 9.04\,pc & C57 & 8.94\,pc & S57 & 10.44\,pc \\
    M90 & 9.46\,pc & C90 & 7.54\,pc & S90 & 11.38\,pc \\
    M57U & 10.95\,pc & C57U & 10.98\,pc & S57U & 14.41\,pc \\
    M57O & 8.22\,pc & C57O & 8.21\,pc & S57O & 9.21\,pc \\
    \hline
\end{tabular}
\label{table:pointspread}
\end{table}


\subsection{Probability distribution functions}

PDFs are an invaluable tool in probing the complex processes that affect cloud evolution --- gravitational contraction, magnetic field effects \citep{molina2012density,federrath2013star,hennebelle2019role}, stellar feedback \citep{krumholz2005general, vazquez2010molecular}, and Mach number of turbulence \citep[e.g.][]{vazquez1994hierarchical,federrath2008density,konstandin2012new,molina2012density}. The current understanding of molecular cloud PDFs is that low density regions, tracing turbulent material, give rise to a log-normal distribution while high density regions, characterised by gravitationally unstable, free-falling material and star-forming cores, give rise to a power-law distribution \citep{burkhart2018star}.

Figure~\hyperref[fig:PDFs]{7a} shows the mass-weighted PDF for all simulations at the final epoch. The solenoidal seeded models show a significantly larger fraction of diffuse regions, followed by mixed and compressive seeded turbulence. To support this and further de-clutter the figure, we plot the cumulative distribution function (CDF) and group different turbulent seeds and different virial ratios for comparison in Figure~\hyperref[fig:PDFs]{7b,c,d,e}. 
As the cloud evolves between the two epochs being compared, the increase in diffuse regions is manifested in the CDF as a shift upwards for $n<100\cmc$ across all simulations. 

In Figure~\hyperref[fig:PDFs]{7b,c}, diffuse regions dominate solenoidal seeded clouds. The different modes bring a very clear separation in the distribution during the early stages of the cloud, which is further deepened as the cloud evolves. Compressive seeded clouds are characterised by fewer low density regions while mixed seeded clouds lies between the two extremes. At the high density end, the difference between turbulent modes is unnotable due to local dynamics taking over the decayed turbulence.

In Figure~\hyperref[fig:PDFs]{7d,e}, comparing initial virial ratios, solenoidal seeded models already show a lag in their evolution at early stages. Across all modes of turbulence, we see a common trend at the low-density end: underbound $>$ virial balanced $>$ overbound, with the latter two following each other closely. S57U again remains an extreme case, showing a very large fraction of low density regions. We interpret this as the combination of the solenoidal turbulence mode and the initial underbound state of the cloud both impeding the local and global collapse, resulting in an excess of diffuse regions as compared to the other clouds. At the high density end, which is dominated by the strength of local (and to some extent global) gravitational collapse, we see dense regions preferentially forming in overbound clouds (see Figure~\hyperref[fig:PDFs]{7a}).

To test these claims more rigorously, we perform Kolmogorov-Smirnov (KS) tests on the CDFs for all simulations with the same RNG seed$=57$. The $p$-values for each KS test are shown in Figure \ref{fig:KStests}. Comparing the bottom-left, bottom-center and center-left `tiles', we find that the initial gravitational potential of the cloud has a weak effect on the lower end of the density distributions. This is apparent in the distributions for the underbound clouds, which are most likely to differ from the initial virial equilibrium and overbound cloud distributions.

It is clear, however, that the initial turbulence modes have a much stronger effect on the lower end of the density distributions. The most pronounced difference in the underlying distribution as measured by the KS test can be seen for the solenoidal underbound (S57U) distribution compared to all other CDFs. We also find a stark difference between C57O and all the other \textit{underbound} cases, with the $p$-values indicating that the underlying CDF for C57O is different from the other underbound distributions up to almost 3$\sigma$ for C57U and beyond for M57U and S57U, again suggesting a combined effect between the turbulence mode and initial virial ratio.

Thus, the low density end of distributions is governed by a decreased efficiency of collapse which can be brought about by two mechanisms: having dominant solenoidal modes of turbulence, or by a less gravitationally bound cloud. The effect of either is the same, and in combination can give an extremely diffuse cloud. We conclude that this low density regime, dominated by turbulence, maintains a \textit{strong relic of the initial turbulent mode} and is weakly sensitive to the initial gravitational potential of the cloud.


\subsection{Sink mass}
\label{ssec:sink mass}

\subsubsection{Mass growth}
\label{sssec:mass evolution}

We now look at the evolution of sink mass in the cloud. Figure \ref{fig:sink_mass_evolution} shows that compressive seeded turbulence is characterised by a very early onset of star formation, followed by mixed and lastly solenoidal seeded turbulence. In fact, by the time solenoidal seeded clouds reach $250\msun$ in sinks, the compressive clouds are $\approx 10$ times higher in their stellar mass. This is consistent with previous studies that note similar star formation suppression in solenoidal clouds \citep[][]{federrath2012star,padoan2014star}. The mixed turbulence seeded clouds lag by about $0.5\myr$ ($\approx 7.5\%$ of free-fall time $t_\text{ff}$) compared to their compressive counterparts; in turn, the solenoidal seeded clouds lag by almost $1\myr$ ($\approx 15\% \text{ of } t_\text{ff}$) to the mixed modes. The compressive seeded clouds also have a higher accretion rate onto sinks when comparing each RNG seed. In general, the effect of RNG seed on the total sink mass \textit{at a given timestep} is very large, with variations of $>50\%$ at 0.5 $t_\text{ff}$, akin to the findings of \citet{jaffa2022chaotic}.

Despite the mixed and compressive seeded underbound clouds commencing star formation earlier than their overbound counterparts, the rate of accretion is higher for overbound clouds and they catch up around the second epoch, when $M_\text{sink}=500\msun$. The same is true for the solenoidal case, although S57U shows a remarkably delayed collapse. \citet{howard2016simulating} studied variation in boundedness for a larger $\alpha_\text{vir}$ range and 10 times more massive molecular cloud, and found a similar delay in cluster formation but higher star formation rates for overbound clouds. Their work also suggests, although at larger scales than ours, that initial gravitational boundedness has greater effect than radiative feedback during the first $5\myrs$, with the latter only affecting star-formation efficiency by a few percent.

\subsubsection{Mass distribution and brown dwarf mass excess}
\label{sssec:mass distribution}

We now look at the distribution of the sinks masses, i.e. the sink mass function, for the first and final epoch in Figure~\ref{fig:Sink_mass_stats}. Note that this is truncated to only the star-forming regime $>0.1\msun$. An understanding of the origin of stellar IMF is crucial for the microphysics of galactic evolution and formation of planetary systems, and has been the subject of a great deal of numerical simulations. Some of the physical processes studied in this context include fragmentation \citep[e.g.][]{zinnecker1984star,larson1985cloud,klessen1998fragmentation}, turbulence \citep{padoan2002stellar}, accretion \citep[e.g.][]{bonnell1997accretion,bonnell2001accretion,bate2005origin}, mergers \citep[e.g.][]{bonnell2002accretion}, or combination of many processes \citep{adams1996theory}. See \citet{offner2014origin} for a review on stellar IMF studies.

At the first epoch, the number and distribution of sinks is seemingly independent of the turbulent mode. The median sink mass lies somewhere in the $0.3\rangeto0.5\msun$ range, and and is consistently higher for overbound clouds than underbound clouds which have a more bottom-heavy distribution. This effect is less pronounced in the compressive case (C57U) where the competing effects of convergent flow and diffuse collapse balance each other. The solenoidal underbound cloud (S57U) shows an interesting feature --- the cloud has an abundance of sink particles, roughly $\times3$ more than the other simulations, but the distribution is dominated by low-mass sinks. The variation between the turbulence modes in virial balance is not significant enough just yet.

At the final epoch, the initial distributions are maintained at the lower end by continual production of low mass sinks and at the higher end by the accretion of mass onto heavy sinks. As the clouds evolve, solenoidal seeded turbulence shows a characteristic abundance of sinks, dominating in number, but with a bottom-heavy distribution so that each individual sink mass is very low. The trend seen for S57U at the first epoch is now also seen across the other solenoidal modes at the final epoch. 

To compare distributions, now also including sinks below the $0.1\msun$ threshold, we plot the mass function of the sinks in Figure~\ref{fig:Sink_mass_histogarm}. As suggested by Figure~\ref{fig:Sink_mass_stats}, the characteristic mass for underbound clouds is lower than overbound clouds. Furthermore, across all simulations, we see that the mass functions roughly follow the \citet{chabrier2005initial} IMF model but is steeper at the low mass end, and shallower at the high mass end (the exact IMF shape is sensitive to binning at high masses. The shape and characteristic masses also vary largely with the turbulent seed. Most interestingly, an abundance of sinks in the solenoidal seeded clouds are of sub-stellar mass, with distributions peaking below 0.1\,M$_\odot$ --- equating to an \textit{excess at brown dwarf masses}! This leads us to question --- \textit{where are these low mass sinks forming}? To investigate this, we plot the spatial distribution of sinks, split into low mass ($<0.1\msun$) and high mass ($>0.1\msun$) mass, in S26 in Figure~\ref{fig:BrownDwarfDistribution}. The two mass regimes show similar spatial clustering, which is also seen in the other solenoidal simulations. This suggests that the brown dwarf mass sinks are not formed in unique regions that prevent accretion.

Mixed and compressive clouds attain massive sinks by efficient accretion and have a larger fraction of sinks at the high mass end, but dont show any noticeable difference between the two turbulent modes. At the low mass end, underbound clouds produce more low mass sinks compared to their overbound counterparts, while high mass sinks are favoured by the overbound clouds. This is in agreement with \citet{howard2016simulating} who see the same effect in cluster formation for a $10^6\msun$ cloud. In the case of S57O, the competing effects of the less efficient solenoidal turbulence and compressive gravitational energy counteract each other, producing a mass distribution similar to other mixed and compressive seeded simulations. 

Figure~\ref{fig:Massive_sink_count} shows the number of sinks which are candidates for massive star formation, with masses greater than 16\,M$_\odot$ (yielding >8\,M$_\odot$ stars assuming 50\% efficiency). The number of massive sinks is comparable for all simulations, with the exception of S57U as it is an extreme case, meaning that the most massive clumps form regardless of initial conditions. However, the relative fraction of massive sinks is greater for overbound clouds (except in case of C57O) compared to underbound clouds, and mixed/compressive seeded clouds compared to solenoidal seeded clouds, seen in both the mass function (Figure~\ref{fig:Sink_mass_histogarm}) and count histogram (Figure~\ref{fig:Massive_sink_count}). The abundance of sinks in the solenoidal seeded clouds likely means that they are able to coalesce some together, making it possible to have a comparable number of massive sinks.

We do not include any radiative feedback in our simulations, instead choosing to focus only on the first few$\myrs$ of collapse. The presence of protostellar heating could potentially change the abundance of brown dwarfs \citep[see, e.g.,][]{bate2009importance} or affect sink-formation in the immediate vicinity of massive protostars. Barring this caveat, we arrive at a similar result to the PDF investigation, where the low mass end of the distribution is controlled by the presence of turbulent solenoidal modes in the cloud or low gravitational potential energy. At the high mass end, the distribution is dominated by strong gravitational collapse and mixed/compressive turbulent seeds.

\subsubsection{Observational aspects}

On the side of observations, studies of the Central Molecular Zone \citep[CMZ,][]{federrath2016link,rani2022solenoidal} and the Orion B cloud complex \citep{orkisz2017turbulence} have found that the low star formation efficiency in these clouds is likely due to stronger solenoidal modes in turbulence. The CMZ, being at the Galactic centre, is certainly subject to a high shear environment exciting these solenoidal modes which decrease in fraction with Galactocentric distance. Our results here are consistent with this idea of star formation inhibition.

The presence of moderately large numbers of low-mass brown dwarfs in clusters like Trapezium, IC 348 and $\rho$ Oph \citep{luhman2000initial} could perhaps be explained by the solenoidal mechanism. Although these clusters have not been studied in this context, the velocity PDFs of the Huygens region in Orion Nebula (home to the Trapezium cluster) show characteristics of solenoidal turbulence \citep{anorve2019analysis}. Furthermore, studies of spatial distribution of sub-stellar objects in Taurus \citep{luhman2006spatial,parker2011mass}, $\rho$ Oph \citep{parker2012search} and IC 348 \citep{parker2017dynamical} reveal that they are not clustered differently than higher mass stars, consistent with our simulations and hint at a common formation mechanism across the mass range.


\begin{figure}
\centering
    \includegraphics[width=1.0\linewidth]{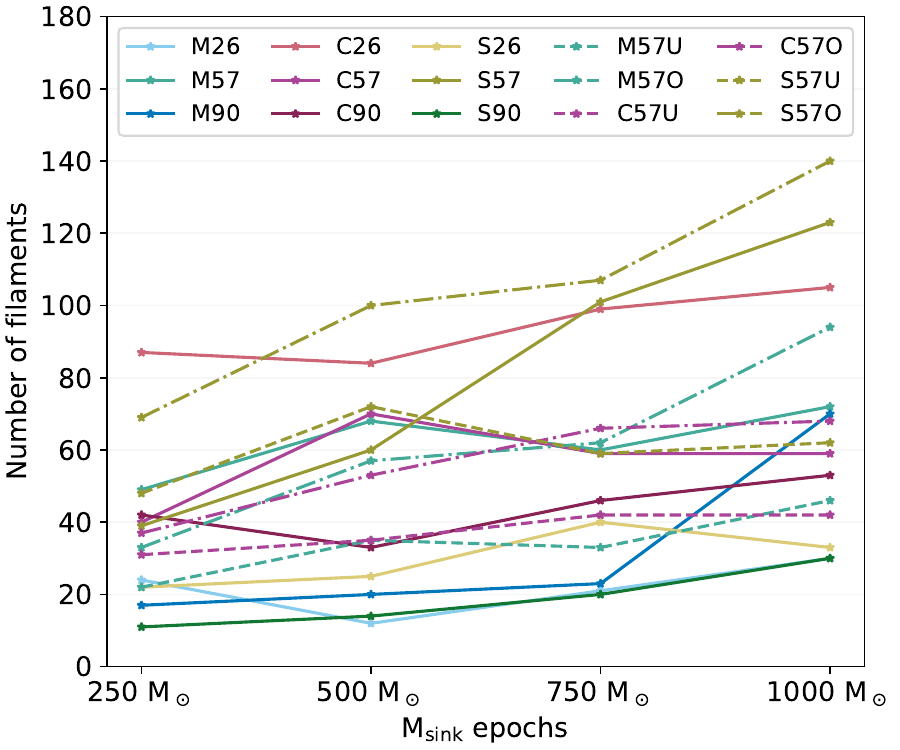}
    \caption{Number of filaments at the four epochs of interest for all simulations, showing a consistent or increasing trend.}
    \label{fig:Filament_number_evolution}
\end{figure}

\begin{figure*}
    \begin{subfigure}{1.0\linewidth}
        \centering
        \includegraphics[width=0.95\linewidth]{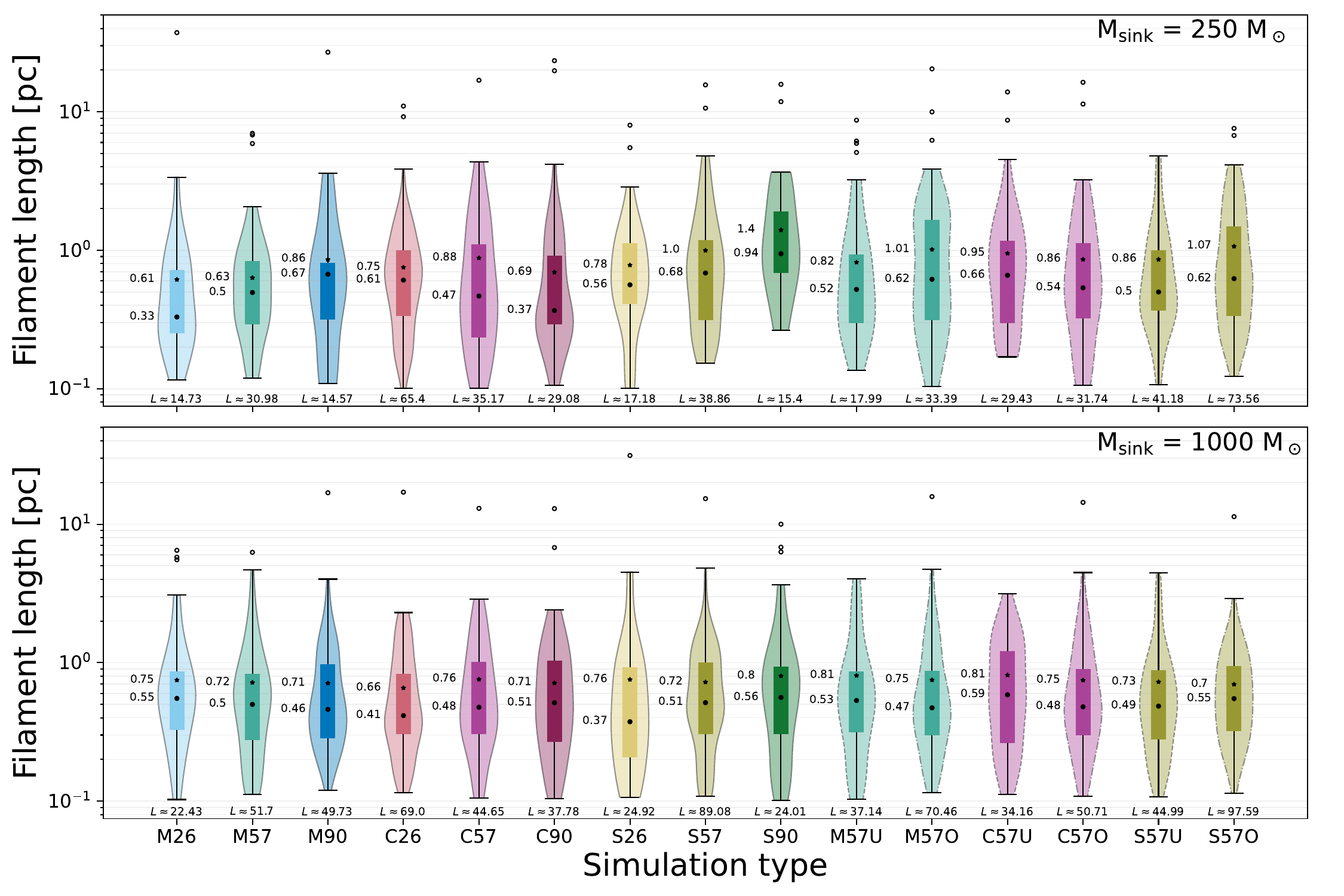}
        \subcaption{Violin plots showing the filament length statistics at 
        $M_\text{sink}=250\msun$ and $M_\text{sink}=1000\msun$ epochs. The plots show 25\%–75\% quartiles (coloured bars), along with the median (black dot) and mean (black star). Empty circles are outlier filaments $>5\pc$.}
        \label{fig:Filament_length_stats}
    \end{subfigure}
    \begin{subfigure}{1.0\linewidth}
        \centering
        \includegraphics[width=0.95\linewidth]{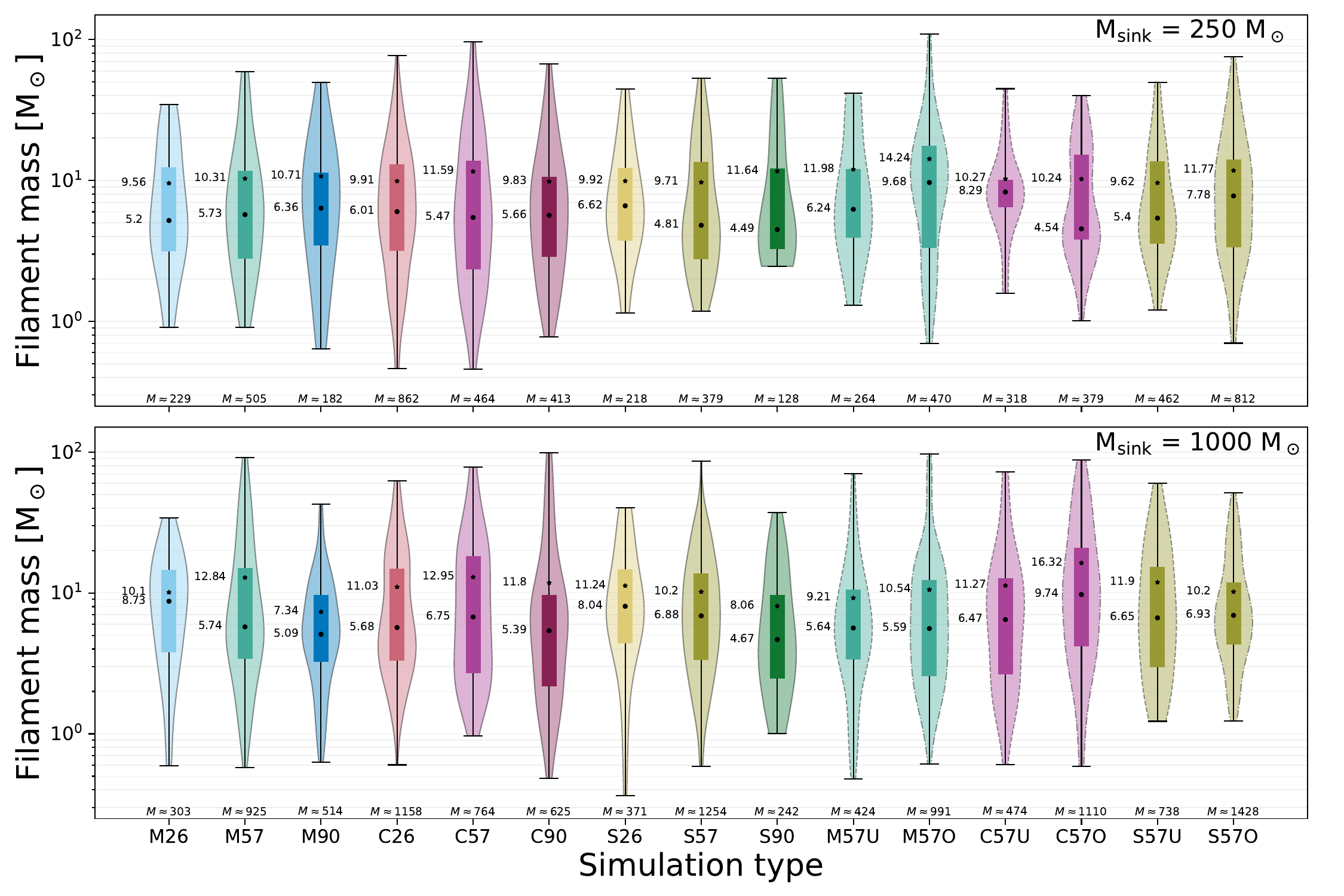}
        \subcaption{Same as \textbf{(a)} but for filament mass.}
        \label{fig:Filament_mass_stats}
    \end{subfigure}
\end{figure*}
\begin{figure*}\ContinuedFloat
    \begin{subfigure}{1.0\linewidth}
        \centering
        \includegraphics[width=0.95\linewidth]
        {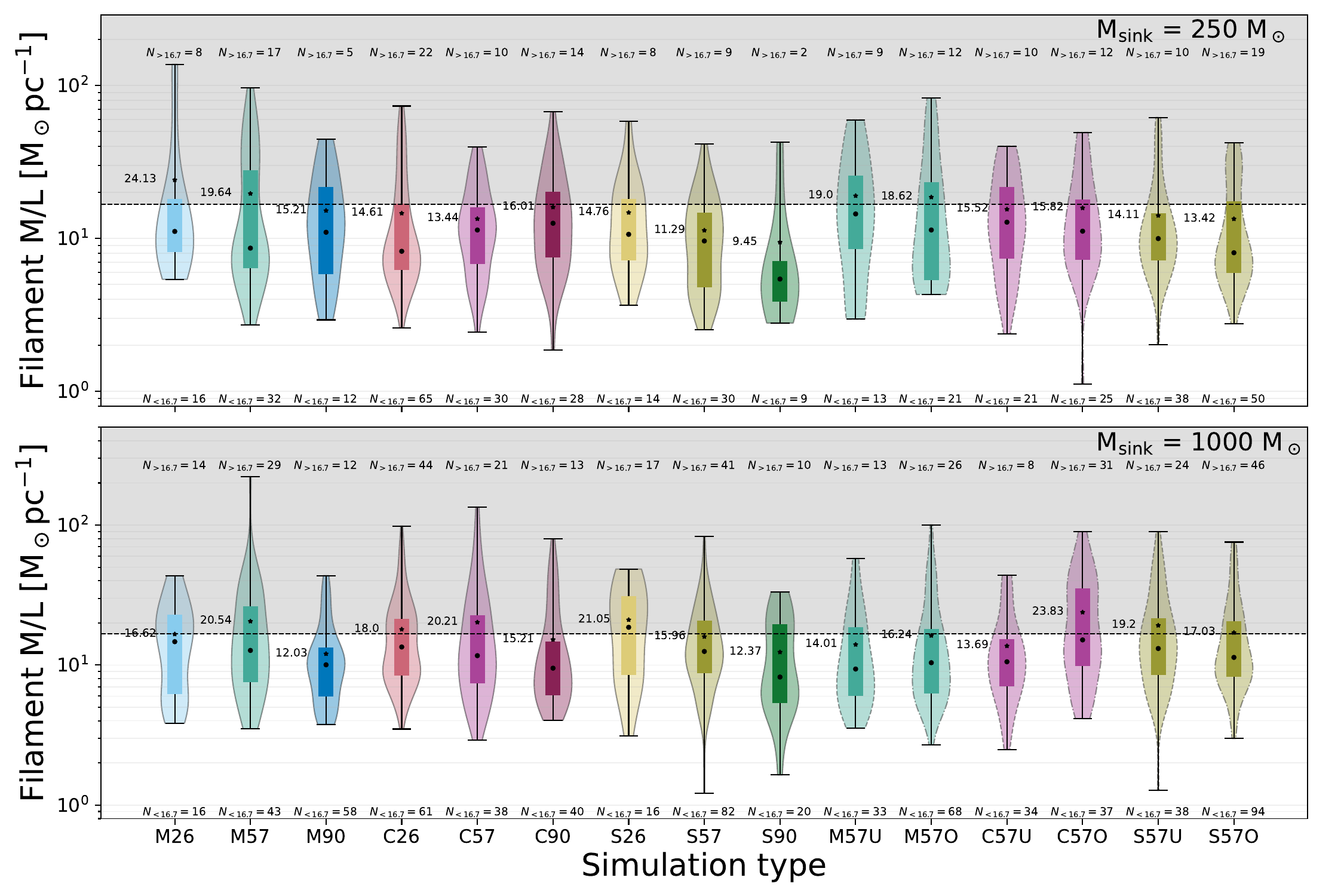}
        \subcaption{Same as \textbf{(a)} but for filament line density. The $M/L$ space is divided into sub-critical (white) and super-critical (grey) regimes, with $M_\text{line,cr} \approx 16.7 \msun\text{pc}^{-1}$ (dotted black line). $N_{<16.7}$ and $N_{>16.7}$ indicate the count of filaments in the sub and super-critical regimes respectively.}
        \label{fig:Filament_ML_stats}  
    \end{subfigure}
    \caption{}
\end{figure*}

\begin{figure}
\centering
    \includegraphics[width=1.0\linewidth]{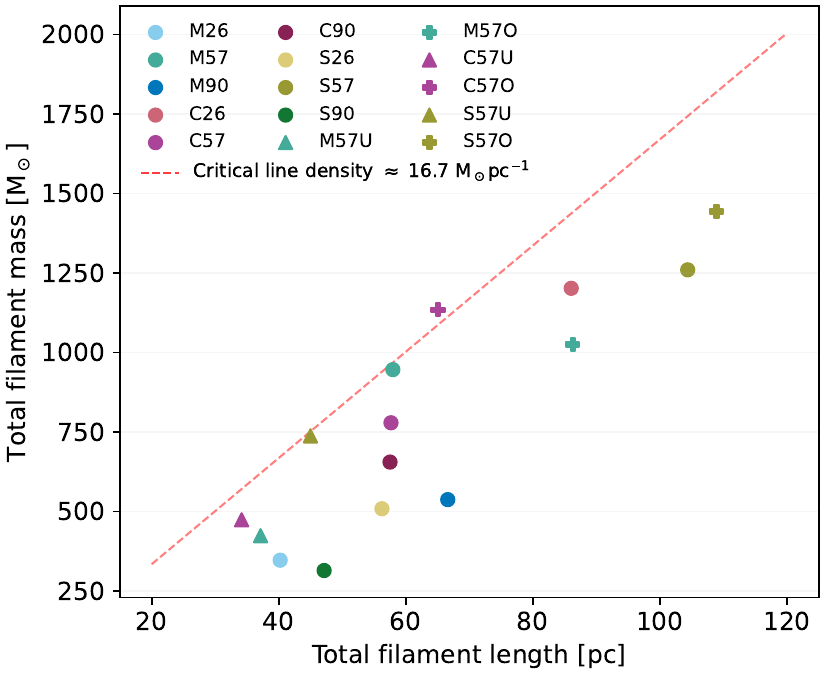}
    \caption{Total mass versus total length of filamentary network at the final epoch $M_\text{sink}=1000\msun$ for all 15 simulations. As number of filaments increase, so does the total length of the network and proportionally the total mass.}
    \label{fig:Filament_mass_vs_length}
\end{figure}

\subsection{Filament networks}

We now look at the nature of filamentary structures produced in the simulations, identified and characterised using the methodology detailed in Section~\ref{ssec:Filament identification} and \ref{ssec:fiesta}. The total number of filaments at each epoch of interest is shown in Figure~\ref{fig:Filament_number_evolution}. Over time, there is a marginal increase in the number of filaments in most simulations, however some show greater variability such as M90, S57 and S57O. Furthermore, although overbound and virial balanced clouds don't show significant differences, they produce more filaments than their underbound counterparts.

\subsubsection{Filament length}

Figure~\ref{fig:Filament_length_stats} shows the distribution of filament lengths in the simulations. The lower limit of $0.1\pc$ is artificially set by the resolution for \disperse, and tenuous filaments of length $>5\pc$ are shown as outliers. Remarkably, there is very little differentiation between the initial turbulent mode or virial ratio in the distributions. There is a marginal evolution over time towards shorter lengths which perhaps suggests fragmentation. One global distinction comes from the total length of the network in the last epoch which follows the trend: overbound $>$ virial balanced $>$ underbound, likely because the overbound and virial balanced clouds produce more filaments. Figure~\ref{fig:Filament_mass_vs_length} highlights this trend clearly. Overall, the last epoch shows an interesting consistency in median length of $\approx0.5\pc$, and mean length of $\approx0.7\rangeto0.8\pc$ across all simulations.

\subsubsection{Filament mass}

Figure~\ref{fig:Filament_mass_stats} shows the distribution of filament masses in the simulations. As with the length distributions, there is no clear distinction between the initial turbulent modes, and no clear evolution in their statistics, other than accretion as all simulations gather more mass onto filaments. The mean filament mass of $\approx10\msun$ and median of $\approx 5\rangeto6\msun$ remains consistent across simulations and time. The trend of overbound $>$ virial balanced $>$ underbound seen in the lengths at the final epoch is also reflected in the masses, so they are roughly proportional, as seen in Figure~\ref{fig:Filament_mass_vs_length}. In fact, filaments in overbound clouds carry twice as much mass in total as underbound clouds. Simulations with more high density regions in their PDFs (Figure~\ref{fig:PDFs}a), such as C26, M57, S57, M57O, S57O and C57O, also favour more mass in filaments.

\subsubsection{Filament line density}

We now look at how the mass per unit length evolves in the clouds. Note that the mass-to-length ratio here is defined as $M/L$ where $M$ is the total mass of the filament and $L$ is its total length; hence, the quantity referred to here is the average mass-to-length ratio rather than an integrated quantity. Early works \citep{ostriker1964equilibrium} used idealised filament geometry to study fragmentation into cores/clumps, suggesting a critical line density of 
\begin{equation}
    \label{eqn:Mlinecr}
    M_\text{line,cr} = 2 c_s/G \approx 16.7\left(\dfrac{T}{10\kelvin}\right)\msun\pc^{-1}.
\end{equation}
As prefaced in the introduction, and examined in detail by the authors in \citet{chira2018fragmentation}, this simplistic model does not account for the complexity of fragmentation, with sub-critical filaments able to fragment and super-critical filaments seen in both observations and simulations. Figure~\ref{fig:Filament_ML_stats} shows the distribution of filament line densities seen in our simulations. We see a few extreme super-critical filaments, but a majority of them ($\sim50\%\rangeto75\%$) sit just below the critical line. This is also seen in Figure~\ref{fig:Filament_mass_vs_length} where almost all filament networks (on average, in their entirety) are subcritical with respect to thermal pressure. Since we are looking at fixed epochs rather than evolving time-steps, it is likely that the extreme cases have very low survivability, and may be too fragmented to identify as a filaments. As seen with the filament masses, overbound clouds form more super-critical filaments than underbound clouds. There are marginally more super-critical filaments in compressive clouds in the first epoch, but this is washed out in the final epoch, and the only correlation that remains is those simulations producing more high density regions having more super-critical filaments. As the cloud evolves, we see the expected increase in number of filaments in the super-critical regime.

\subsubsection{Comparison with previous work}

Statistical analysis of filaments in 3D is sparse in literature. The closest work to ours is that of \citet{smith2014nature,smith2016nature}. In \citet{smith2016nature}, the authors identify sub-filaments in a cloud with very similar initial conditions and mixed turbulence, and characterize velocities within them. After $4\myrs$, they find a mean sub-filament mass of $\sim10\msun$ consistent with our work. The longer filaments ($\gtrsim1.5\pc$) identified in \citet{smith2014nature} with masses of $40\rangeto100\msun$ are roughly in line with the higher end of simulations M26, M57 and M90 seen in Figures~\ref{fig:Filament_length_stats} and \ref{fig:Filament_mass_stats}. \citet{clarke2017filamentary} study accretion-driven turbulence in a $3\pc$ cylinder and find greater heirarchical and filamentary substructure in mixed turbulence compared to compressive; although we do not study hierarchy specifically, we find no evidence for it in the number of filaments. \citet{chira2018fragmentation} simulate a much larger $40\,\mathrm{kpc}$ long vertical box with a minimum $1.7\pc$ filament length, beyond the length scale relevant to this work.

A direct comparison to observations proves difficult since they identify filaments in 2D column density maps. \citet{zamora2017fibres} investigated 3D filaments in clouds with similar initial cloud mass, and found that these fibres (the filament family for $\lesssim 1\pc$ length scales) in one 2D projection are not necessarily fibres in other projections. \citet{suri2019carma} use position-position-velocity (PPV) cubes to study such fibres/short filaments in 3D in Orion A, but focus on filament widths which we fix in our work.


\subsection{Hubs}

\begin{figure}
\centering
    \includegraphics[width=1.0\linewidth]{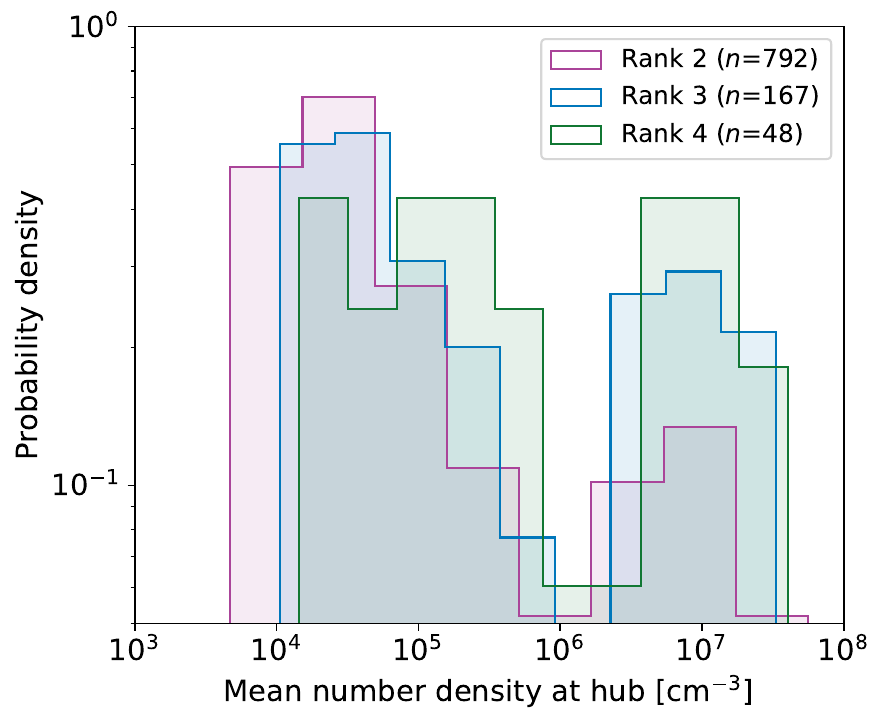}
    \caption{Mean number density at hubs (in a sphere of radius $r=0.05\pc$), for different hub ranks. It can be seen that higher rank hubs exist in denser regions.}
    \label{fig:HubDensity}
\end{figure}

\begin{figure*}
\centering
    \includegraphics[width=0.75\linewidth]{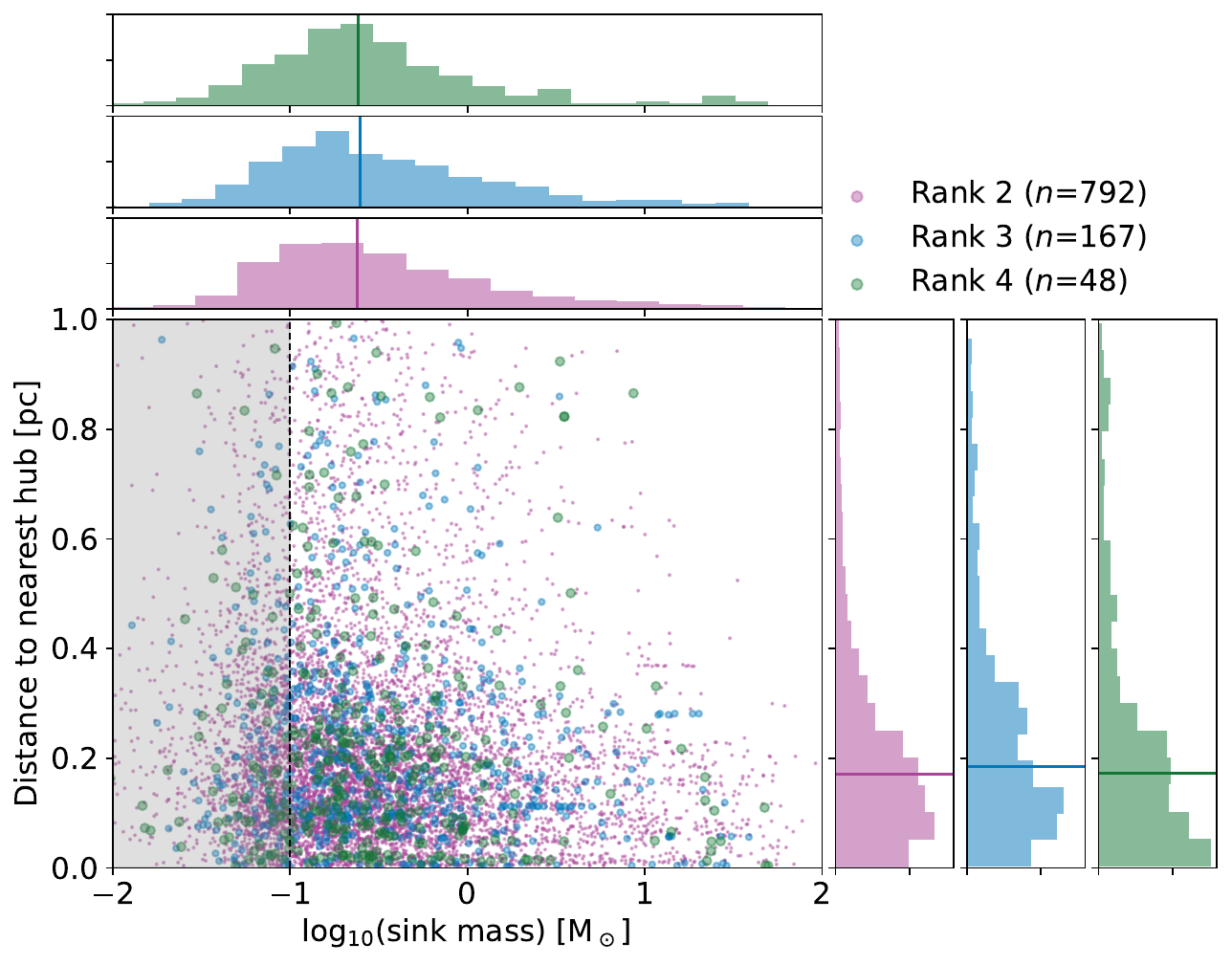}\caption{Distance of individual sinks from the closest hub as a function of their sink mass for all simulations at $M_\text{sink}=1000\msun$, with different colors indicating the rank of the closest hub. The distribution along $x$ and $y$ axes for each hub rank are plotted as histograms with median (solid line). The grey region demarcates sink masses below $0.1\msun$, which are unlikely to form stars.}
    \label{fig:HubSinkRelation}
\end{figure*}

As discussed in the introduction, there exists a possible link between massive-star formation and ‘hubs’, regions where distinct filaments meet. We present a simple model here to characterise them in our simulations: hubs are identified in a filamentary network by constructing the set of all points in filaments, and checking the number of repetitions (if any) of each point in different filaments, which we define as its \textit{rank}. Rank-$n$ points, for $n>1$, are classified as hubs where $n$ filaments converge.

To study these hubs, we pick the $M_\text{sink}=1000~\msun$ epoch since the sink particles have had more time to accrete mass and form potential sites of massive stars. Figure~\ref{fig:HubDensity} shows the mean number density in a radius $r=0.05\pc$ around a hub, for different hub ranks across all simulations. All hubs lie above $n>1000\cmc$, as expected from dense regions. Since rank-2 hubs are just where two filaments meet, they are far more abundant than rank-3 and rank-4 hubs. Rank-4 hubs are primarily high density spots, followed by rank-3 and rank-2 hubs which suggests that the geometry of these junctions is indeed correlated to their density. A possible scenario is that as dense regions within a filament fragment, they create shorter filaments and naturally a junction between them. There also exists an interesting bi-modality in the distribution, which is less prominent prior to filament spine correction (see Figure~\ref{fig:HubDensity_uncorr} in Appendix~\ref{appendix2}), although the correlation between hub rank and density still remains. It is possible that longer filaments, just above the \text{DisPerSE} identification threshold, produce the first mode whereas shorter, fragmented filaments around dense clumps produce the second mode.

We also investigate, for the first time on these scales, the link between massive star formation and the hubs. Figure~\ref{fig:HubSinkRelation} shows the distance to the nearest hub for all sinks, as a function of sink mass across the simulations. 
Low mass sinks, in the range $0.01\msun\rangeto1\msun$, show some preference for hub activity but are highly scattered. They are more likely to be strongly affected by dynamics of their surroundings, being potentially slung away from hubs. High mass sinks ($>1\msun$) show less scatter and a preference for hub activity. Hub ranks, despite being correlated to local density, have very similar distribution in the sink masses they host and their distance to these sinks.

Recent observations of hub-filament systems are limited to high-mass, supercritical filaments/large cores \citep[several hundred$\msun$, e.g.][]{williams2018gravity,kumar2020unifying,kumar2022filament}. As more of these systems are found on lower mass scales, it will shed light on how protostars interact with these dense hubs, and compare to simulations.


\section{Caveats and future work}
\label{s:caveats}

Although care has been taken to select a reasonable persistence threshold for filament identification in \disperse, the properties of filaments will always depend on the method of identification, and the particular set of parameters for that method -- do filaments that bend at angles $>90^\circ$ count as two merging filaments or one? How does one account for discontinuities in filaments? An irregular clump or just a short filament? Applying the same methodology across simulations (such as fixing the filament identification parameters and filament width in our case) provides a means for comparative work, but cannot produce definitive conclusions. 

Furthermore, our simulations also omit important feedback processes such as supernovae explosions, stellar winds, jets and outflows. These would become crucial in the late-stage evolution of the clouds, hence why our focus lies in the early stages to reduce this effect. We also lack magnetic fields, which may play a role particularly in the formation of filaments. This is an issue that would be worth exploring in future work.

Our simulations and Python library have opened up the possibility for a lot more to be studied. The statistics of the turbulent velocity field and its evolution on local and global scales are particularly interesting, and the relation between velocity flows and densities along the filament could reveal a lot about accretion mechanisms.


\section{Conclusions}
\label{s:conclusions}

In this paper, we study the effects of large-scale turbulence in molecular clouds and its links to dense filamentary networks and star-forming regions. We perform 15 high-resolution \arepo\, simulations of a spherical, molecular cloud of mass $10^4\msun$ and radius $\approx9\pc$, seeded with different types of decaying turbulence. Nine of these simulations focus on comparing mixed, purely compressive and purely solenoidal modes of turbulence in virial balance, while the other six focus on comparing different virial ratios, producing gravitationally overbound and underbound clouds. The suite of simulations is publicly available on \texttt{Zenodo}. Dense cores and clumps are formed in the simulations, and sink particles are used to characterise star-formation. Filaments are identified using \disperse, and characterised using novel algorithms released as a Python toolkit called \fiesta. Our main conclusions from the simulations are as follows:

\begin{enumerate}
	\item Clouds seeded with different initial turbulent modes collapse differently, and a different timescale on which star-formation occurs. Compressive seeded turbulence results in an early onset of collapse into stars, approximately $1.5\myrs\rangeto2\myrs$ after the initial collapse ($30\%$ of 1 free-fall timescale $t_\text{ff}$), followed by mixed turbulence. Solenoidal seeded turbulence is the slowest, with star-formation beginning only $3\myrs\rangeto4\myrs$ ($60\%$ of $t_\text{ff}$) after collapse. The mass accretion rate is higher for compressive modes, and also for overbound clouds.
	\item Compressive seeded and overbound clouds give rise to a centrally condensed morphology, with more compact global structure, while solenoidal seeded and underbound clouds give rise to more filament-like morphologies. The latter is characterised by a more diffuse cloud, spread over a larger volume.
	\item The low mass end of the gas density distribution (PDFs) maintains an imprint of the initial turbulent mode. Solenoidal seeded turbulence produces more low-density regions, as do underbound clouds. However, the high-mass end of the PDF is dominated by gravitational collapse (both local and global) and hence is favoured in overbound clouds.
	\item The sink mass functions (representing the mass going into stellar systems) are dominated by brown dwarf mass sinks in the case of solenoidal seeded clouds. These clouds generate an abundance of sinks, far more than  the other turbulent modes, but are indeed bottom-heavy in their mass distribution. The high-mass regime generally constitute a larger number fraction in mixed/compressive seeded and overbound clouds, but the most massive sinks ($>16\msun$) form equally across all simulations.
	\item The distribution of filament lengths and masses is seemingly unaffected by the mode of turbulence seeded, since they trace dense regions where the initial turbulence is decayed and potentially washed over by local collapse dynamics. However, clouds with higher gravitational potential energy produce more filaments, and hence longer networks and more total mass in filaments. Across all simulations, the filaments accrete mass and fragment over time, increasing the abundance of super-critical filaments. In the later stages of collapse (at $M_\text{sink}=1000\msun$ in our simulations), the mean filament length is $\approx 0.75\pc$ while the mean mass is $\approx10\msun$. 
	\item Hubs/junctions where filaments converge show proportionally higher densities with increasing number of converging filaments. Sinks have a preference for hub activity, with low mass sinks showing larger scatter away from hubs. The distribution of sinks is agnostic to the number of filaments converging at the hub.
\end{enumerate}

To succinctly summarise the role of initial conditions in molecular clouds: the \textit{initial turbulence imprints diffuse regions} in the clouds, seen through the different diffuse morphologies, brown dwarf mass excess and low density end of the PDFs; while initial gravity dominates dense regions, seen through the high density end of PDFs, abundance of mass in filaments, and increased hub activity.

\section*{Acknowledgements}

JD and ZF would like to thank Kamran Bogue, David Whitworth and Jiancheng Feng, for their informative discussions during the course of the Masters project at University of Manchester. They especially thank RJS for the opportunity, support and mentorship during the project which have made this paper possible. RJS gratefully acknowledges an STFC Ernest Rutherford fellowship (grant ST/N00485X/1) and HPC from the Durham DiRAC supercomputing facility (grants ST/P002293/1, ST/R002371/1, ST/S002502/1, and ST/R000832/1) without which this work would not have been possible. JD, ZF and RJS would additionally like to thank Sansith Hewapathirana for bug-testing and confirming some of the results obtained from the \arepo\, simulations.

This research made use of \textsc{astropy}, a community-developed python package for Astronomy \citep{robitaille2013astropy,price2018astropy}, \textsc{matplotlib} \citep{hunter2007matplotlib}, \textsc{numpy} \citep{oliphant2006guide} and \textsc{scipy} \citep{virtanen2020scipy}.

{\section*{Data Availability}
\phantomsection
\label{sec:Data Availability}}

The relevant \arepo\, snapshots and \disperse\, filament networks for all 15 simulations studied in this work are publicly available at \href{https://doi.org/10.5281/zenodo.7946648}{10.5281/zenodo.7946648}. Furthermore, the Python library developed for analysing the simulation results and characterising filaments, \fiesta, is available at \href{https://fiesta-astro.readthedocs.io}{fiesta-astro.readthedocs.io}. The initial turbulent velocity fields, due to their large file sizes, are available on request.
\newline



\bibliographystyle{mnras}
\bibliography{library} 


\appendix

\section{Persistence threshold analysis and error estimation}
\label{appendix1}

\begin{figure*}
\centering
    \includegraphics[width=1.0\linewidth]{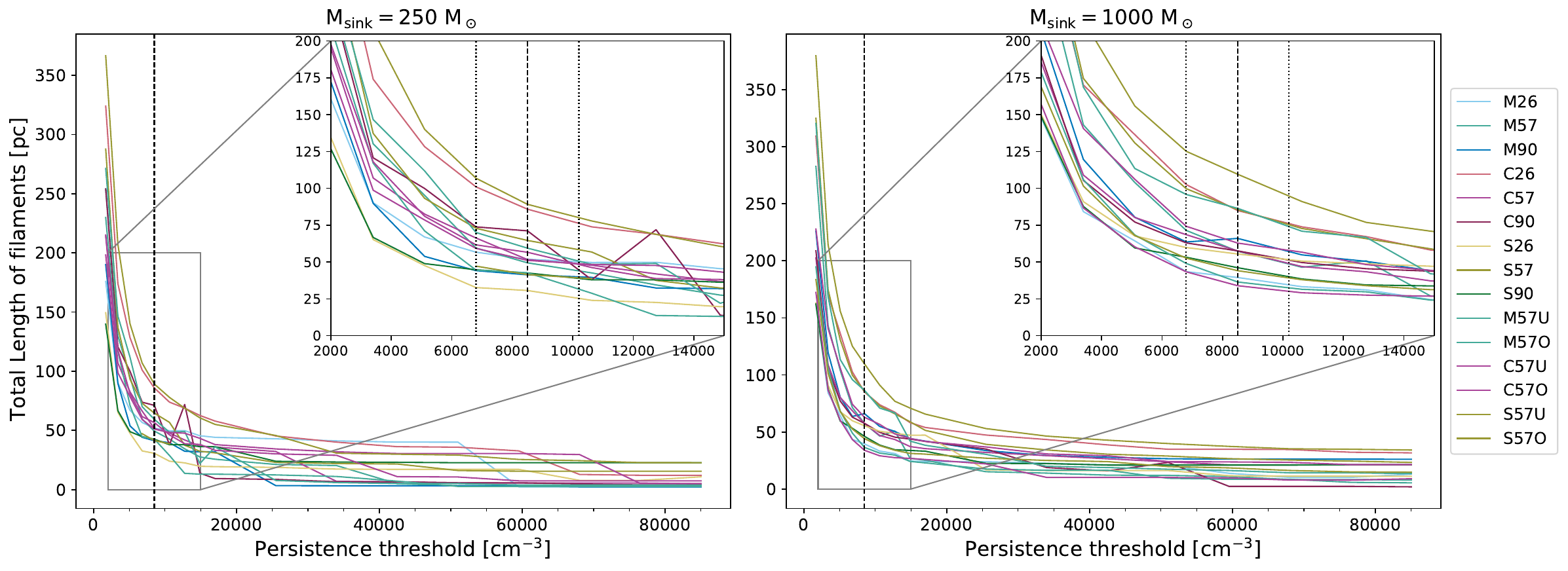}
    \caption{Total length of filaments as a function of \disperse\, persistence threshold for the two epochs $M_\text{sink}=250\msun$ and $M_\text{sink}=1000\msun$. The dashed line indicates the ideal persistence threshold, used for analysis in this paper, while the dotted lines indicate $3\sigma$ confidence region around this value.}
    \label{fig:PersistenceThreshold}
\end{figure*}

We discuss here the methodology used to determine an ideal persistence threshold for identifying filaments using \disperse\, in our simulations. The mean number density of all simulation clouds lies around $n\approx 100\cmc$, as seen in Figure~\ref{fig:PDFs}. Since we're only interested in dense filamentary structures that lie well over the average molecular cloud densities, we vary the persistence threshold in the range $850\cmc$ to $85000\cmc$. In order to investigate how the filament skeleton varies with choice of threshold, we pick the total length of filaments as our metric. This is a fairly robust measure since the same filament skeleton can be fragmented to inflate the number of filaments and consequently decrease the average length, while still maintaining the same total length and overall structure. Furthermore, any metric that uses the filament mass or density will depend on the assumption determining the extent of a filament, and the algorithm used to query \arepo\, cells around the filament. Figure~\ref{fig:PersistenceThreshold} shows the total length of all filament networks at epochs $M_\text{sink}=250\msun$ and $M_\text{sink}=1000\msun$ for a range of persistence thresholds. High thresholds provide very few, if any, filaments. Viewing the filamentary structures at these thresholds revealed that only long, tenuous filaments roughly tracing dense regions were being identified. Low thresholds lead to a rapid increase in the total length; this is due to the spurious identifications of a high number of very short filaments.

Zooming into the region at the elbow of the plots, we test various persistence thresholds and view the resultant filament skeletons in 3D and 2D against the simulations. To see the local number densities traced by these filaments, we mask regions with $n<5000\cmc$, as done in Figure~\ref{fig:FilamentExample}, and find $n_\text{ideal} = 8500\cmc$as the ideal persistence threshold that traces regions $n>5000\cmc$ well, and define a $3\sigma$ confidence region around this value. The errors calculated in this manner are quantified in Table~\ref{table:PersistenceError} for global filament properties in simulations M57, C57 and S57. The effect of persistence threshold parameter is of-course non-linear and this is only used as a metric to see the magnitude of errors, which vary from $<1\%$ to $20\%+$ in the worst case.


\begin{table}
\centering
\caption{Example of $1\sigma$ errors introduced in filament properties at the final epoch due to persistence threshold parameter in \disperse.}
\def\arraystretch{1.7}
\begin{tabular}{c|c|c|c}
    \hline
    \hline
    	& M57 & C57 & S57 \\
    \hline
    $n_\text{fil}$ & $72^{+5.0}_{-5.3}$ & $59^{+4.7}_{-5.0}$ & $123^{+10.0}_{-5.0}$\\
    \hline
    $L_\text{mean}$ & $0.72^{+0.013}_{-0.000}$ & $0.76^{+0.007}_{-0.000}$ & $0.72^{+0.000}_{-0.016}$ \\
    $L_\text{median}$ & $0.50^{+0.001}_{-0.003}$ & $0.48^{+0.024}_{-0.016}$ & $0.51^{+0.000}_{-0.014}$ \\
    \hline
    $M_\text{mean}$ & $12.84^{+0.442}_{-0.152}$ & $12.95^{+0.845}_{-0.427}$ & $10.20^{+0.080}_{-0.355}$ \\
    $M_\text{median}$ & $5.75^{+0.359}_{-0.000}$ & $6.75^{+1.712}_{-0.484}$ & $6.88^{+0.393}_{-0.414}$ \\
    \hline
\end{tabular}
\label{table:PersistenceError}
\end{table}

\section{Effect of filament corrections on hubs}
\label{appendix2}

\begin{figure}
\centering
    \includegraphics[width=1.0\linewidth]{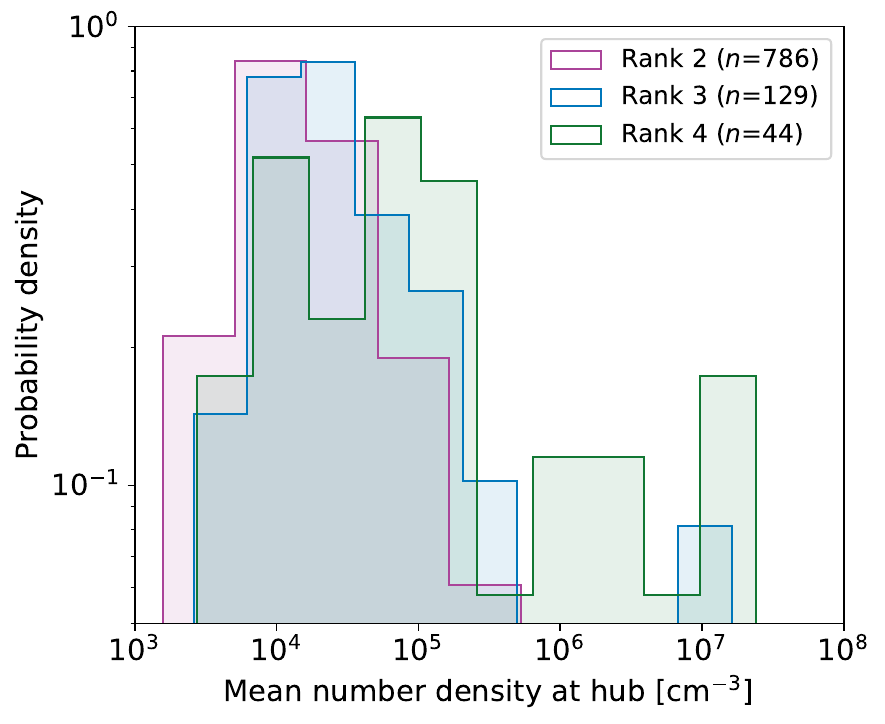}
    \caption{Mean number density at hubs (in a sphere of radius $r=0.05\pc$), for different hub ranks \textit{before} correcting filament spines (c.f. Figure~\ref{fig:HubDensity}).}
    \label{fig:HubDensity_uncorr}
\end{figure}

Figure~\ref{fig:HubDensity_uncorr} shows the plot of mean number density in a radius $r=0.05\pc$ around a
hub for different hub ranks, similar to Figure~\ref{fig:HubDensity}, but \textit{before} the filament spines are corrected. The filament correction inherently depends on using local \arepo\, density to reduce the error in \disperse's identification of filaments on a coarsely resolved grid (as detailed in Section~\ref{sss:caveatstocharacterization}). This naturally affects the density of hubs, and thus the aforementioned Figures, but the main conclusion of a strong correlation between hub rank and density remains. Even in the uncorrected filaments, rank-4 hubs form the densest regions followed by rank-3 and rank-2. The clumpy nature and complex hierarchy of structure within filaments adds difficulty in rigorously defining hubs and making simple assertions regarding their properties.
\vspace{20em}

\bsp    
\label{lastpage}
\end{document}